\documentclass[aps,prl,twocolumn,superscriptaddress, showpacs]{revtex4}
\pdfoutput=1
\usepackage{epsfig}
\usepackage{amssymb}
\usepackage{bm}
\usepackage{graphicx,color}
\usepackage{subfig}
\usepackage{graphicx}
\usepackage{mwe}
\usepackage{dcolumn}
\usepackage{epstopdf}
\usepackage{csquotes}
\usepackage{amsmath}
\usepackage{tikz,xcolor,hyperref}
\definecolor{lime}{HTML}{A6CE39}
\DeclareRobustCommand{\orcidicon}{
	\begin{tikzpicture}
	\draw[lime, fill=lime] (0,0) 
	circle [radius=0.16] 
	node[white] {{\fontfamily{qag}\selectfont \tiny ID}};
	\draw[white, fill=white] (-0.0625,0.095) 
	circle [radius=0.007];
	\end{tikzpicture}
	\hspace{-2mm}
}
\foreach \x in {A, ..., Z}{\expandafter\xdef\csname orcid\x\endcsname{\noexpand\href{https://orcid.org/\csname orcidauthor\x\endcsname}
			{\noexpand\orcidicon}}
}

\begin{document}



\title{First Measurement of Neutron Birefringence in Polarized $^{129}$Xe and $^{131}$Xe Nuclei}
\author{H. Lu\orcidB{}}\affiliation{Indiana University/CEEM, 2401 Milo B. Sampson Lane, Bloomington, IN 47408, USA}
\author{M.J. Barlow} \affiliation{School of Medicine, University of Nottingham, Queens Medical Centre, Nottingham, UK}
\author{D. Basler} \affiliation{School of Chemical and Biomolecular Sciences, Southern Illinois University, Carbondale, IL 62901, USA}
\author{P. Gutfreund} \affiliation{Instut Laue-Langevin,  71 Avenue des Martyrs, CS 20156, 38042 Grenoble Cedex 9, France}
\author{O. Holderer} \affiliation{Forschungszentrum Jülich GmbH, Jülich Centre for Neutron Science (JCNS) at Heinz Maier-Leibnitz Zentrum (MLZ), 85747 Garching, Germany}
\author{A. Ioffe} \affiliation{Forschungszentrum Jülich GmbH, Jülich Centre for Neutron Science (JCNS) at Heinz Maier-Leibnitz Zentrum (MLZ), 85747 Garching, Germany}
\author{S. Pasini} \affiliation{Forschungszentrum Jülich GmbH, Jülich Centre for Neutron Science (JCNS) at Heinz Maier-Leibnitz Zentrum (MLZ), 85747 Garching, Germany}
\author{P. Pistel} \affiliation{Forschungzentrum J\"ulich mbH, ZEA-1, 52425 J\"ulich Germany}
\author{Z. Salhi} \affiliation{Forschungszentrum Jülich GmbH, Jülich Centre for Neutron Science (JCNS) at Heinz Maier-Leibnitz Zentrum (MLZ), 85747 Garching, Germany}
\author{K. Zhernenkov} \affiliation{Forschungszentrum Jülich GmbH, Jülich Centre for Neutron Science (JCNS) at Heinz Maier-Leibnitz Zentrum (MLZ), 85747 Garching, Germany}
\author{B.M. Goodson\orcidC{}} \affiliation{School of Chemical and Biomolecular Sciences, Southern Illinois University, Carbondale, IL 62901, USA}
\author{W.M. Snow\orcidD{}} \affiliation{Indiana University/CEEM, 2401 Milo B. Sampson Lane, Bloomington, IN 47408, USA}
\author{E. Babcock\orcidA{}} \affiliation{Forschungszentrum Jülich GmbH, Jülich Centre for Neutron Science (JCNS) at Heinz Maier-Leibnitz Zentrum (MLZ), 85747 Garching, Germany} \email[]{e.babcock@fz-juelich.de}

\date{\today}
\begin{abstract}

We present the first measurements of polarized neutron birefringence in transmission through nuclear-polarized $^{129}$Xe and $^{131}$Xe gas and determine the neutron incoherent scattering lengths $b_i(^{129}Xe)=0.186\pm(0.021)_{stat.}\pm(0.004)_{syst.}\space\text{ fm}$ and $b_i(^{131}Xe)=2.09\pm(0.29)_{stat.}\pm(0.12)_{syst.}\space\text{ fm}$ for the first time. These results determine the essential parameter needed for interpretation of spin-dependent neutron-scattering studies on polarized xenon ensembles, with possible future applications ranging from tests of time-reversal violation to mode-entangled neutron scattering experiments on nuclear-polarized systems. 

 \end{abstract}

\maketitle
This work presents the first measurement of neutron birefringence in polarized $^{129}$Xe and $^{131}$Xe nuclei and the first measurement of the nuclear polarization-dependent bound scattering length difference $\Delta b=b_{+}-b_{-}$ for nuclear spin $I$ parallel or antiparallel to the neutron spin $s$. Knowing $\Delta b$ one can for the first time now conduct and interpret spin-dependent neutron scattering from an ensemble of polarized xenon nuclei using the well-established theory of Van Hove~\cite{VanHove1954, Lovesey2003} generalized for neutron spin-dependent scattering from polarized nuclei~\cite{Schermer1968}. Because nuclear-polarized xenon atom ensembles can be created in conditions where the electron spins do not dominate the magnetic properties (unlike the great majority of magnetic systems in condensed matter), our work enables qualitatively new types of polarized neutron investigations. Highly-polarized ensembles of xenon gas can be created by spin-exchange optical pumping (SEOP)~\cite{WalkerHapper,bascontflow,rosen1999,zook,meersmann129,hersmanPRL,NikolaouhighP2,NikolaouhighP,sheffield}
 in volumes high enough to create long-lived polarized Xe liquids and solids by freezing~\cite{gatzke1993,Sauer1997,Haake1998} for exploration of subtle properties of these ``pure" spin systems. The conclusions of this paper describe examples of possible future polarized neutron investigations which make essential use of the special properties of polarized xenon in quantum entanglement \cite{Katz2022a,Katz2022b} and in searches for new sources of time reversal violation. These newly-enabled neutron scientific applications of polarized $^{129}$\/Xe and $^{131}$\/Xe can also complement their many existing applications in biomedical imaging~\cite{WalkerHapper,boydrev2002,NikolaouhighP2,sheffield,lungrev,Khan2021} (including new MRI/gamma-ray imaging modalities~\cite{meta131nature}), NMR spectroscopy~\cite{boydrev2002,danillarev}, fluid dynamics~\cite{boydrev2002,lungrev,danillarev}, gas/surface interactions~\cite{Wu1987,Wu1988,Raftery1991}, studies of Berry geometric phases~\cite{mehringgyro}, and searches for  CPT/Lorentz violation~\cite{bear2000,cane2004,gemmel2010,Allmendinger2014,Stadnik2015,Kostelecky2018}, electric dipole moments~\cite{Sachdeva2019}, and axion-like particles~\cite{walkerNsnow, Feng:2022}. 
 
Neutron scattering amplitudes are often expressed in operator form as $b=b_{c}+b_{i}\vec{s} \cdot \vec{I}$ where $\vec{s}$ is the neutron spin, $b_{c}={[(I+1)b_{+}+Ib_{-}]}/{(2I+1)}$ is the spin-independent coherent scattering length, and the spin-dependent incoherent scattering length $b_{i}={I\sqrt{I+1}[b_{+}-b_{-}]}/{(2I+1)}$ is directly proportional to $\Delta b$. 
For $^{129}$Xe or $^3$He with nuclear spin $I={1/2}$ the compound neutron-nucleus total spin $J=I\pm s$ can form a triplet, ($J=1$) and singlet ($J=0$) total spin state corresponding to the $b_1\equiv b_+$ and $b_0\equiv b_-$ channels, so $\Delta b=b_1-b_0$. For $I={3/2}$ $^{131}$Xe, $m_{J}=2,1,0,-1$ are possible. $\Delta b$ measures the difference of the $b_{m_{J}=2}+b_{m_{J}=1}$ and $b_{m_{J}=0}+b_{{m_{J}=-1}}$ scattering amplitudes for spin order characterized by the nuclear polarization $P_x={<I_z>/I}$ with no tensor alignment. A general analysis of neutron spin dynamics in media with nuclear spin order~\cite{Gudkov20} implies that, for the precision reached here and for the neutron energies far from neutron-nucleus resonances used in this work, we can relate the spin rotation angle to the scattering length difference in the usual way. Texts on neutron optics~\cite{Sears} discuss the statistical weight factors used to derive the above relations.

We measured $\Delta b$ by observing the precession of the neutron spin as neutrons pass through a polarized nuclear target, named ``pseudomagnetic precession"~\cite{Baryshevsky1965} in the literature. Although this phenomenon was initially described~\cite{Baryshevsky1965, Abragam1982} in terms of a fictitious ``pseudomagnetic field" inside the sample, $\Delta b$ originates from neutron-nucleus scattering. The optical theorem~\cite{Sears} relates the spin dependence of the neutron optical potentials associated with the scattering amplitudes $b_{+}$ and $b_{-}$ to a two-valued neutron index of refraction ($n_{+}$,$n_{-}$)  depending on the relative orientation of the neutron spin and the nuclear polarization:
\begin{equation}
    \begin{split}
        n^{2}_{\pm}=1-{\frac{4\pi}{k^{2}}}N(b_{coh}+b_{\pm}), \\ \\
\Delta n=(n_+-n_-)\approx -{\frac{2\pi}{k^{2}}}N (b_{+}-b_{-}),
\label{eq:newindex2}  
      \end{split}
\end{equation}
\noindent where $N$ is the number of nuclei per unit volume, $k$ is the neutron wave number, and the approximation in the second expression is valid in our case as the neutron index of refraction is $\simeq 1$. $\Delta n$ makes the medium optically birefringent for neutrons so that the two helicity components of the neutron spin state accumulate different phases, $kn_{\pm} d$, in the forward direction as neutrons propagate a distance $d$ through the target. Therefore neutron spins orthogonal to the nuclear polarization direction of the target precess around the nuclear polarization by an angle $\phi^*=k \Delta n d$. 

\begin{figure}
\begin{center}
\includegraphics[width=20pc]{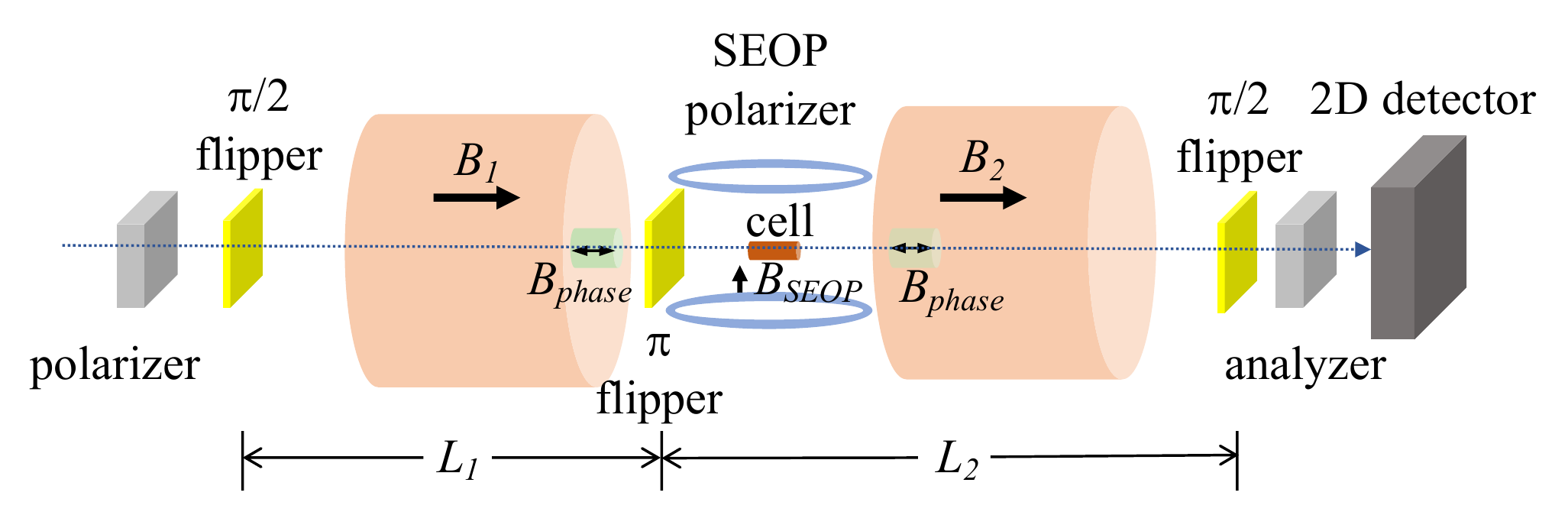}
\end{center}
\caption{\label{NSEdiagram}A schematic drawing of an NSE instrument following the description in the text. The first neutron $\pi/2$ flipper sets the neutron spins to precess about the total field defined by the direction of $B_1$, $B_{SEOP}$, or $B_2$, $i.e.$ in the respective regions. The $\pi$ flipper reverses neutron precession and the second $\pi/2$ flipper and analyser return the in-phase magnitude of the final neutron polarization.}
\end{figure}

Measurements of neutron birefringence are well-suited to the Ramsey method of separated oscillatory fields~\cite{Ramsey1956, Ramsey1990}. 
Previous work~\cite{Abragam1973, Forte1973, Pokazanev1979, Abragam1982, Tsulaia2014} determined $\Delta b$ for several nuclei dynamically-polarized in the solid state. We used the neutron spin-echo technique (NSE)~\cite{Mezei1972} to measure $\Delta b$ in SEOP cells filled with $^{3}$He, $^{129}$Xe, or $^{131}$Xe (an earlier measurement in $^3$He~\cite{Zimmer2002} also used this method). The measurement sequence is similar to spin echo manipulations in nuclear magnetic resonance~\cite{Hahn1950}, however the precession and flipping fields are encountered in space along the traveling neutron beam, as opposed to time-dependent fields applied to spins at rest in the lab frame. In contrast to the Ramsey sequence, NSE uses a $\pi$ spin flip at the field symmetry point (Fig.~\ref{NSEdiagram}) to refocus the spin precession of neutrons with different velocities so they are rephased at the polarization analyzer. Phase shifts of the interference fringes from the sample are compensated by DC magnetic fields from phase (compensation) coils which are scanned over several periods about the compensation point to obtain the NSE signal. The sensitivity of the measurement is therefore set by the ratio of the field resolution in the compensation coils to the total field integral of the instrument. Since for the J-NSE instrument~\cite{neutrons} used in this work the phase coil precision can be nT/m compared to a total field integral on the instrument of over 1 T/m, very high phase precision is possible.

The spin-echo condition holds for any group of neutron velocities at $B_{1,echo}=\frac{L_{2}}{L_{1}}B_{2}$ where the number of forward precessions though field $B_1$ over length $L_1$ in the first region and back precessions in the second region of field $B_2$ and length $L_2$ are equal, $i.e.$~$\phi_{1}(B_{1,echo}) = \phi_{2}$. The phase shift accumulated in either region is $\frac{\gamma m \lambda}{2 \pi \hbar}B_1L_1$ where m is the mass of a neutron with de Broglie wavelength $\lambda$ and gyromagnetic ratio $\gamma$. The additional phase shift from $\Delta b$ modifies the spin echo condition by adding an extra phase $\phi^*$. The precession caused by the neutron birefringence is 
\begin{equation}
    \label{eq:pseudoangle1/2}
    \begin{split}
\phi^{*} &=-{\frac{2I}{2I+1}}\lambda P_xN_xd_x{\Delta b^x}\\
\\
&= -2\sqrt{\frac{I_x}{I_x+1}} \lambda P_xN_xd_xb_{i}^x.
\end{split}
\end{equation}

Here $I_{3}=I_{129}=1/2$ for $^{3}$He and $^{129}$Xe  and $I_{131}=3/2$ for $^{131}$Xe, and $P_x$ and $N_x$ are the polarization and number density of the respective polarized nuclei of atomic weight $x$ with corresponding scattering length difference $\Delta b_i^x$ or incoherent scattering length $b_i^x$. The relevant product $P_xN_x$ is determined by NMR calibration measurements using absolute $P_3N_3$ of the $^{3}$He cell from neutron transmission as a standard.  $\phi^{*}$ is then measured by the shift of the NSE signal upon reversal of the nuclear polarization, with all static magnetic fields constant. The NSE signal $i(B_{phase})$ is the transmitted intensity after the neutron polarization analyzer as a function of the current in the solenoid phase coil field $B_{phase}$. 

Fig.~\ref{NSEdiagram} shows the self-compensated superconducting (SC) coil sets for the two precession regions of the J-NSE~\cite{Pasini2019}. The polarized noble gas samples rest in the sample region inside a $B_0$ holding field normal to the neutron beam. Three GE180 SEOP cells~\cite{Rich2001} produced in FZ-J\"ulich were used in this experiment: a 5 cm inner diameter, 4.9 cm long $^{3}$He cell, and two Xe cells, both 5 cm inner diameter and 12.7 cm long.  All cells contained several mg of Rb, the $^{3}$He cell had $0.3556$ bar of pure $^{3}$He gas with $0.1$ bar of $N_{2}$, the $^{129}$Xe cell was filled to $0.3$ bar of 91.18\% enrichment $^{129}$Xe gas with $0.25$ bar of $N_{2}$, and the $^{131}$Xe cell was filled with $0.20$ bar 84.4\% enriched $^{131}$Xe (Berry and Associates / Icon Isotopes) and 0.2 bar $N_{2}$.  The $^{3}$He and $^{129}$Xe cells were prepared in Garching (FZ-J\"ulich)~\cite{Salhi2014} and the $^{131}$Xe cell was prepared and characterized at Southern Illinois University~\cite{Boyd1,Boyd}. 

Two frequency-narrowed diode array bars~\cite{Babcock2016} realized in-situ SEOP. A 70 cm diameter Helmholtz coil pair produced the magnetic field $B_{SEOP}$ normal to the neutron path. Cell heating and temperature regulation was provided by AC electric cartridge heaters for the $^3$He cell and by flowing air for the two xenon cells. 

Nuclear magnetic resonance free-induction decay measurements of the cell magnetizations directly proportional to $P_xN_x$ used a home-built pulse-receive NMR system~\cite{Babcock2016} with a single 2 cm diameter, 300-turn pickup coil placed on the cell's surface normal to both the optical pumping axis and the neutron beam. Since the three isotopes studied here vary in gyromagnetic ratios by an order of magnitude ($\gamma_3=-3.243$ kHz/g, $\gamma_{129}=-1.178$ kHz/g, and $\gamma_{131}=0.349$ kHz/g for $^3$He, $^{129}$Xe, and $^{131}$Xe respectively), the NMR FID calibrations presented an experimental challenge.

\begin{figure}
\begin{center}
\includegraphics[width=20pc]{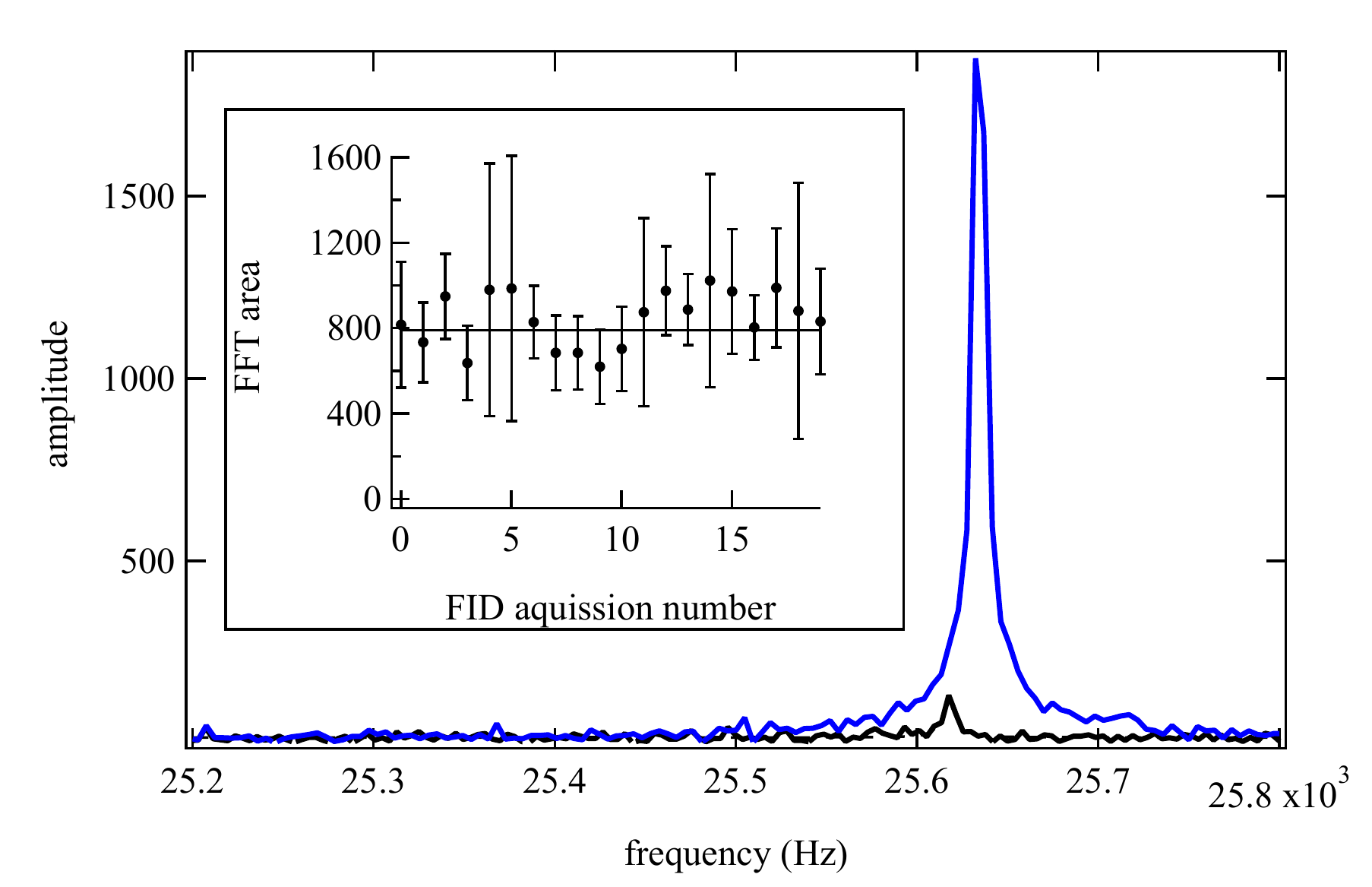}
\end{center}
\caption{\label{131XeNMR}Example NMR spectra from Fourier-transformed single-shot FID signals recorded during the calibration of the $^{131}$Xe cell. $^{131}$Xe and $^{129}$Xe spectra are black and blue, respectively. The inset shows the NMR FID strengths versus acquisition number used for the averaging of the $^{131}$Xe signal.}
\end{figure}

The $^{129}$Xe cell magnetization was measured using NMR which was calibrated to an absolute $^3$He standard from neutron transmission during the NSE measurements. The ratio $R=T_p/T_0$ of the neutron transmission through the polarized $^3$He cell ($T_p$) to unpolarized ($T_0$) determines cosh$^{-1}(R)=\frac{\sigma_{p}}{\lambda_{th}} \lambda P_3N_3d_3$ where $\sigma_{p}=(1-\sigma_1/\sigma_{un})\sigma_{un}$ is the polarized $^3$He spin dependent neutron absorption cross section, $\sigma_{un}$ the total unpolarized neutron absorption cross section, $\lambda_{th}=1.798$ \rm\AA\  is the standard thermal neutron wavelength, $\lambda=8\pm0.08$ \rm\AA\  was the neutron wavelength of the measurement, $P_3N_3$ is the product of $^3$He polarization and density, and $d_3$ is the cell length. We use $\sigma_p\simeq\sigma_{un}=5333\pm7$ barn as done for other $^3$He neutron spin filter cells~\cite{Gentile2017}. $\sigma_p$ is smaller than $\sigma_{un}$ barn by $\sigma_1$, which has been estimated to be $24$ barn~\cite{Huber2014, HuberPC}, leading to a small systematic in $P_3N_3$ on the order of -0.45\%.  Our measurements found $R=1.2506\pm 0.0030$ giving  cosh$^{-1}R=0.6939\pm 0.0040$, which gives $P_3N_3=(0.609\pm 0.145) \times 10^{24}~{\rm m^{-3}}$ or $P_3N_3=0.251\pm0.006$ bar. 
Separate neutron transmission measurements of this cell using neutron time-of-flight~\cite{Chupp2007,Campbell2011} determined $N_3=0.3556 \pm 0.0011$ bar at $298.5$ K in the cell center where $d_3=4.8\pm 0.1$ cm and characterized the shape of its rounded ends. $d_3$ is inferred from measurements of the cell's external length and assumptions of the glass thickness resulting in the given error. This density implies $P_3=70.6 \pm 1.6\%$. Only the product $P_3N_3$ is needed for absolute calibration of our NMR system so the error in $d_3$ is not propagated to the $b_{i}$ results, but the Xe cell lengths are needed to solve for $\Delta b$ of $^{129}$Xe and $^{131}$Xe. The observed $^3$He NMR signal was stable to better than $0.3\%$. 

The neutron wavelength distribution transmitted by the velocity selector is fit by a unity-normalized triangular function convoluted with a cosine function. The resulting form is $i=i_0-A_0\cdot tri({\rm I}_{phase}) \cdot \cos{(\omega({\rm I}_{phase}-{\rm I}_0))}$, where the triangular function $tri$ has a width fixed by the velocity selector, $i$ is the neutron signal intensity, $i_0$ is a constant intensity offset, $A_0$ is the maximum amplitude of the spin echo oscillation (which occurs at the echo condition ${\rm I}_{phase} = {\rm I}_0$), and $\omega$ is the angular frequency of the oscillation. All pixels of the 2D position sensitive neutron detector are analyzed individually and the results averaged. 

The data were taken in defined time-ordered sequences of alternating up and down target polarization for all three nuclei. Since the $^3$He pseudomagnetic precession angle has been measured previously~\cite{Zimmer2002} and was known to be large compared to our expected effects, we used this data to check the experimental procedure and apparatus. 
The up/down polarization states for the polarized xenon targets were switched by reversing the pump-laser polarization by turning the quarter-wave plates without any other changes. The nuclear polarization is reversed by SEOP on a timescale near the $T_1$ relaxation time of $^{129}$Xe, about 5 min. for our cell, so one 20 min NSE scan was skipped after each wave plate change. For $^{131}$Xe, $T_1\simeq 30$ s \cite{Boyd1,Stupic2011} is much shorter than the scan time and the polarization buildup time is negligible. 
    
We analyzed the individual NSE scans in single detector pixels for each run. The spin echo stationary phase point varies slightly across the neutron beam due to the small spatial variations in the field integral, and the spin echo phase drifts slowly over timescales long compared to the target spin flip due to very small changes in the total field over time. To improve the signal/noise ratio for fitting the pixel spin echo scans, we applied a Fast Fourier Transform (FFT) frequency filter to the spin echo data. A frequency bandpass around the main frequency peak having a wide frequency range compared to the distribution of frequencies associated with the fractional neutron wavelength distribution of 10\% was used. To fit the FFT-filtered NSE signals, we first fix the wavelength distribution width and compute the average spin echo frequency $\omega$, which is related to the offset in I$_{phase}$ for each measurement, and was $\omega=35.92, 35.28$, and $35.42$ Hz for $^{129}$Xe, $^{131}$Xe, and $^{3}$He respectively. Then we allow the remaining three parameters, $N_0$, $A_0$, and $\phi_0$ to vary. 
    
\begin{figure}[!ht]
    \includegraphics[width=17pc]{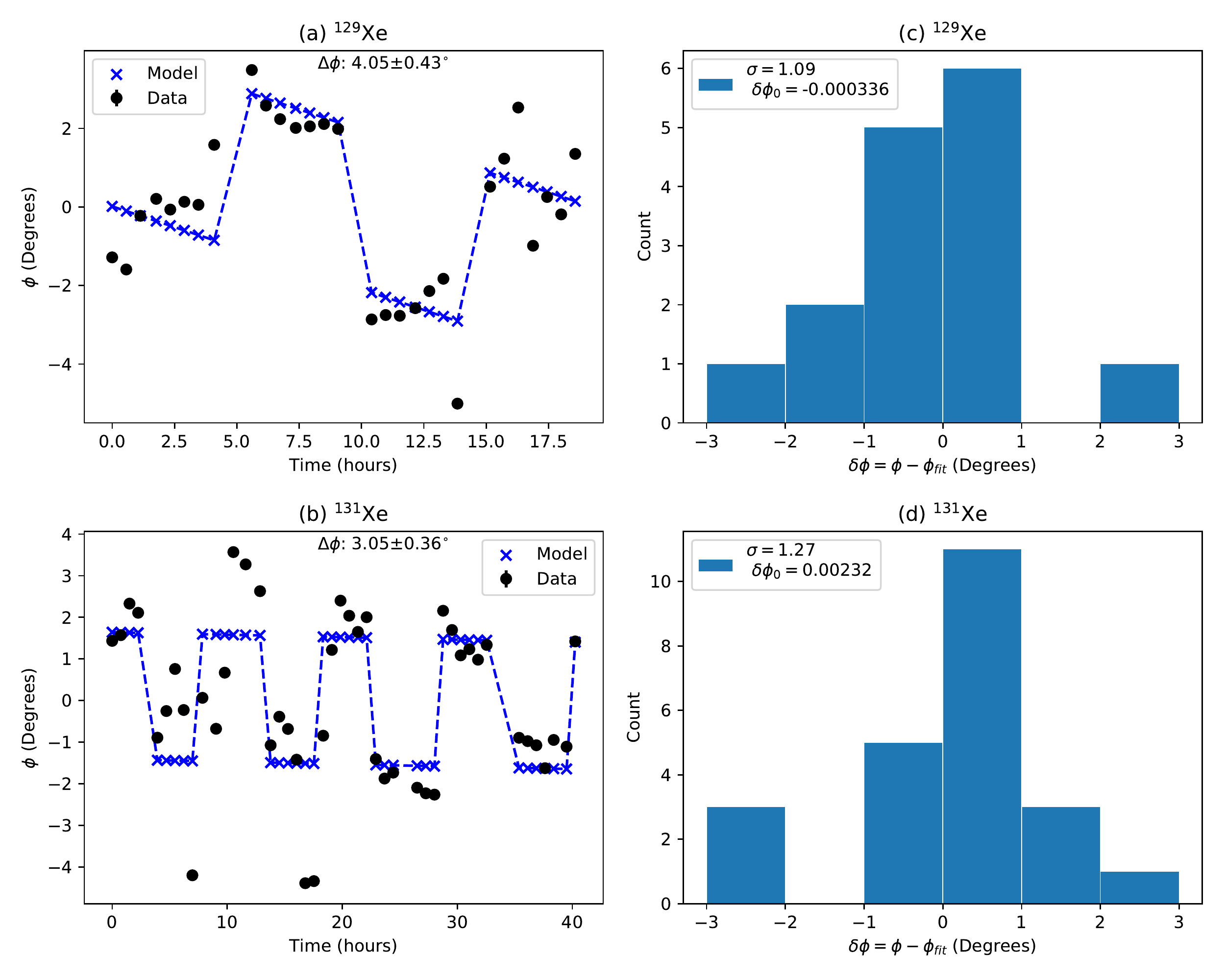}
    \caption{(a,b) Plots of relative NSE phases of $^{129}$Xe and $^{131}$Xe versus experiment time. The data were evenly distributed into groups of alternating $P_{x}$. The fist point after switching of $P_{129}$ is excluded from analysis due to the finite polarization build up time constant, whereas all scans for $^{131}$Xe can be included due to its fast polarization buildup. (c,d) Bar graphs show the distribution of phases about the mean value from the fit.}
    \label{fig:129Xe131Xephis}
\end{figure}

The precession angle is extracted from the relative phase shift between oppositely-polarized nuclear target states with the NSE scans performed using alternating groups of polarizations. The absolute phase $\phi$ was well represented by a square wave on top of a slowly varying linear instrumental phase drift (Fig.~\ref{fig:129Xe131Xephis}). This resulted in precession angles averaged over the active detector pixels of $4.05^{\circ}\pm0.43^{\circ}$ for the $^{129}$Xe target and $3.05^{\circ}\pm0.36^{\circ}$ for the $^{131}$Xe target.

We then use Eq. \ref{eq:pseudoangle1/2} to compute the incoherent scattering lengths $b_i$. Since the NMR calibration of the Xe measures magnetization proportional to $P_xN_x$, any error in $N_x$ for Xe drops out for the determination of $b_i$. Using the NMR calibrations, we determine $P_xN_x$ for the two xenon isotopes as follows. For $I=1/2$ $^{129}$Xe and $^3$He and for NMR FID performed in the very low tip angle limit ($i.e.<<90^\circ$),
\begin{equation}
    \label{eq:PN129from3}
    P_{129}N_{129}=P_3N_3\left({\frac{\gamma_3}{\gamma_{129}}}\right)^2{\frac{S_{129}V_{3}}{S_3V_{129}}},
\end{equation}
where the ratio of gyromagnetic ratios is squared to scale for both the coil pickup and tipping pulse at fixed tip parameters, $S_{x}$ is the NMR strength of the respective noble gas isotope and $V_{129}/V_3=4.085$ was the increase in tip amplitude for $^{129}$Xe to obtain a good signal/noise ratio (S/N). Using this relation $P_{129}N_{129}=1.62\pm 0.04\times 10^{24}$ ${\rm m^{-3}}$. The 0.3 bar total Xe pressure measured during cell filling implies $P_{129}=17.6\%$. 

The $^{131}$Xe NMR calibration could not be performed during the neutron experiment with our standard pickup coil as the lower $^{131}$Xe polarization and the small $^{131}$Xe gyromagnetic ratio lead to very weak signals; also NSE signals could not be obtained at the high $B_0$ field required to obtain the cross-calibration NMR frequency of 25.6 kHz chosen to reach high enough signal to noise ratio. Therefore $^{131}$Xe polarization was calibrated in a separate measurement after the NSE experiment, leaving the SEOP apparatus and conditions unchanged. Using an NMR coil with a 6-fold higher quality factor, the NMR calibration was performed with a maximum $\pi/2$ tip angle for both Xe isotopes, so one factor of the ratio of gyromagnetic ratios drops out of the calibration calculation. 
Additionally one needs to account for the ratio of the different nuclear spins. The relation for the calibration $^{131}$Xe to $^{129}$Xe becomes:
\begin{equation}
    \label{eq:PN131from129}
     P_{131}N_{131}=P_{129}N_{129} \left({\frac{\gamma_{129}}{\gamma_{131}}}\right)\left({\frac{I_{129}}{I_{131}}}\right) {\frac{S'_{131}}{S'_{129}}},
\end{equation}
where $S'_x$ denote the signals obtained for the $\pi/2$ tip angles used for this step. The result is $P_{131}N_{131}=0.0498P_{129}N_{129}=8.1\pm 0.2\times 10^{22}$ m$^{-3}$. Given the 0.20 bar total Xe pressure, $P_{131}=1.96\%$. Neither the specific number densities nor the isotopic concentrations of the xenon isotopes are needed for the neutron scattering length determination with our method using calibrated NMR.

\begin{table}
\begin{tabular}{l c c} 
\hline\hline
  isotope &$^{129}$Xe & $^{131}$Xe \\ [0.5ex] 
 \hline
$\delta b_i$  & \bf{0.112}& \bf{0.150} \\
\hline
{$\delta b_i$ stat.} & \bf{ 0.110} &\bf{ 0.139 } \\
 stat. error source & & \\
 $\delta\phi$ & 0.106 & 0.118\\
 ${\delta/\delta R}$(cosh$^{-1}(R))$ & 0.0057 & 0.0057 \\
 $\delta(S_3/S_{129})$ & 0.028 & 0.028 \\
 $\delta(S'_{129}/S'_{131})$ & & 0.067 \\
 \hline
 $\delta b_i$ syst. &\bf{ 0.023} &\bf{ 0.055} \\
 syst. error source & & \\
 $\sigma_1$ &-0.0045 &-0.0045 \\
 $\delta d_3$ & 0.021 & 0.021 \\
 $\delta d_{129}$ or $\delta d_{131}$ & 0.0079 & 0.0079 \\
 repeatability Xe pol. & & 0.05 \\
 \hline\hline
 \end{tabular}
    \caption{\label{errors}The relative error contributions divided between systematic and statistical sources. The statistical error of the $\phi$ measurement dominates. The estimate of $\sigma_1=24$ barn from~\cite{Huber2014,HuberPC} was used; the - denotes a one-sided systematic that would lower the reported $b_i$ values.} 
\end{table}

We also briefly measured $b_i^3$ to compare with previous results. This measurement is calibrated absolutely from the polarization dependent $^3$He neutron absorption cross section~\cite{Campbell2011}. Our value of $2.280\pm0.020(stat.)+0.015(syst.)$ fm for $b_i^3$ agrees with previous work~\cite{Zimmer2002,Huber2014} and is determined with much higher precision than our $b_i$ values for $^{131}$Xe and $^{129}$Xe. Since $P_3N_3$ is determined by direct measurement of the neutron transmission ratio through a polarized/unpolarized cell for this particular beam, all of the $b_i$ values are independent of the detailed shape or mean value of the neutron wavelength distribution.

Combining Eqs. (\ref{eq:pseudoangle1/2}), (\ref{eq:PN129from3}) and (\ref{eq:PN131from129}), we can write the magnitudes of incoherent scattering lengths for the xenon isotopes in terms of directly-measured experimental quantities as:
\begin{multline}
    b_i^{129}=\frac{\sqrt{3}}{2}\left(\frac{\gamma_{129}}{\gamma_{3}}\right)^2\frac{\phi_{129}}{\cosh^{-1}(R)}\frac{d_{3}}{d_{129}}\frac{S_{3}V_{129}}{S_{129}V_{3}}{\frac{\sigma_{p}}{\lambda_{th}}}\\
\end{multline}
and
\begin{multline}
    b_i^{131}=\frac{3}{2}\sqrt{\frac{5}{3}}\frac{|\gamma_{129}\gamma_{131}|}{\gamma_{3}^2}\frac{\phi_{131}}{\cosh^{-1}{(R)}}\frac{d_{3}}{d_{131}}\frac{S_{3}V_{129}}{S_{129}V_{3}}\frac{S'_{129}}{S'_{131}}{\frac{\sigma_{p}}{\lambda_{th}}}\\
\end{multline}
where ${\cosh^{-1}{(R)}}=0.6939\pm0.0040$ at the center of the $^3$He cell. The last term in the products account for the $^3$He triplet absorption cross section $\sigma_1$ (i.e. total spin of the neutron+$^3$He of 1) compared to the total neutron wavelength dependent unpolarized absorption cross section (this factor is equivalent to Eq. 25 in \cite{Zimmer2002}). We used a previous estimate for the $^3$He triplet cross section of $\sigma_1=24$ barn \cite{Huber2014,HuberPC}. Other possible corrections due to cell geometry or neutron wavelength distribution~\cite{Zimmer2002} are negligible for this work. A soon-to-be-submitted work on $b_i$ for $^3$He~\cite{HaoIP} wll discuss them. 

The values of the incoherent scattering lengths are thus $$b_i^{129}=-0.186\pm(0.021)_{stat.}\pm(0.004)_{syst.}\space\text{fm}$$ and $$b_i^{131}=2.09\pm(0.29)_{stat.}\pm(0.12)_{syst.}\space\text{fm}.$$  Signs of the scattering lengths are determined from the spin directions in the SEOP setup. The statistical errors $10\%$ and $12\%$ for $^{129}$Xe and $^{131}$Xe, respectively, come from the scatter of the phase shift fits shown in Fig.~\ref{fig:129Xe131Xephis}. These values for $b_{i}$ are in line with those of other nuclei. We are not aware of any simple argument that can explain why $|b_i^{131}$ is a factor of 11 larger than $|b_i^{129}$. 

With $b_i(^{129}\rm Xe)$ measured in this work, we could probe the degree of entanglement of polarized $^{129}$Xe spins generated in  atomic collisions in SEOP systems using polarized, mode-entangled neutron beams to measure spin-spin correlation functions as entanglement witnesses for the xenon spin states. A recently-developed quantitative theory for the scattering of mode-entangled neutron beams from spin-correlated dimers~\cite{Irfan2021} can be extended to polarized xenon gas, which can be accurately modeled as an ideal gas with an analytical expression for the neutron dynamic structure factor. SEOP collisions of $I=1/2$ $^{3}$He and $^{129}$Xe atoms with properly-prepared polarized alkali atoms can generate a calculable degree of entanglement in the nuclear spins under certain conditions according to recent work~\cite{Katz2022a,Katz2022b}. The resulting long-lived entanglement in the nuclear spin system is of interest for optical quantum memories~\cite{Dantan2005,Katz2021,Katz2020}. The quantum decoherence of mode-entangled neutron beams passing through dense matter is so small that the measurement of neutron entanglement witnesses for Bell and GHZ inequalities are unaffected~\cite{Shen2020, Lu2020, Kuhn2021}. The transverse spatial separation between the two opposite-spin subbeams created in devices like neutron Wollaston prisms coincides with the range of mean free paths of the polarized $^{129}$Xe gas atoms accessible in SEOP cells. 

Polarized $^{131}$Xe nuclei could be used in a search for new sources of time reversal (T) violation in neutron-nucleus interactions. T violation from some new interaction beyond the Standard Model of particles and interactions is one of the highest intellectual priorities in nuclear/particle/astrophysics, and could shed light on the matter-antimatter asymmetry in the universe according to the Sakharov argument~\cite{Sakharov1967}. The forward scattering amplitude of polarized neutrons in a polarized nuclear target can possess a parity (P)-odd and T-odd term of the form 
$\vec{s_{n}} \cdot (\vec{k_{n}} \times \vec{I})$
where $\vec{s_{n}}$ is the neutron spin, $\vec{k_{n}}$ is the neutron momentum, and $\vec{I}$ is the nuclear polarization. Compound neutron-nucleus resonance reactions are known to greatly amplify parity violation in neutron-nucleus interactions~\cite{Mitchell1999, Mitchell2001}. A 4\% P-odd asymmetry was measured in the $3.2$ eV p-wave resonance in $^{131}$Xe, an amplification compared to nucleon-nucleon P-odd amplitudes of almost $10^{6}$. The theory which successfully predicted this phenomenon long ago~\cite{Sushkov1980, Sushkov1982} implies that P-odd and T-odd interactions between nucleons beyond the Standard Model should also be amplified by a similar factor~\cite{Bunakov1982, Gudkov1991, Fadeev2019}. Neutron transmission measurements involving such coherent neutron-nucleus interactions could provide null tests for time-reversal invariance that are free from contamination by final state interactions~\cite{Bowman2014}. Advances in neutron polarization technology and source brightness added to progress in SEOP polarization of $^{131}$Xe suffice to conduct a sensitive search for axion-like particle (ALP) exchange~\cite{Fadeev2019}, which is poorly constrained by EDM searches for ALP masses above 10 meV~\cite{Dzuba2018}, because here the Standard Model axion relation between axion mass and coupling constant does not apply~\cite{Mantry2014}.

\begin{acknowledgments}
H. Lu and W. M. Snow acknowledge support from US National Science Foundation (NSF) grants PHY-1913789 and PHY-2209481 and the Indiana University Center for Spacetime Symmetries. H. Lu received a Short-Term Grant, 2019 no. 57442045 from DAAD the German Academic Exchange Service. D. Basler and B.M. Goodson acknowledge support from the NSF (CHE-1905341), DoD (W81XWH-15-1-0272, W81XWH2010578), and a Cottrell Scholar SEED Award from Research Corporation for Science Advancement, and thank T. Gafar, C. Kraft, L. Korando, and A. Ruffing for shipment of the SEOP cells. We acknowledge G.M. Schrank for discussions and M. Huber for detailed discussions of NIST work on $b_i^3$ and estimates of $\sigma_1$ for $^3$He.
\end{acknowledgments}
\bibliography{NeutronBirefringence.bib}

\begin{thebibliography}{80}
\expandafter\ifx\csname natexlab\endcsname\relax\def\natexlab#1{#1}\fi
\expandafter\ifx\csname bibnamefont\endcsname\relax
  \def\bibnamefont#1{#1}\fi
\expandafter\ifx\csname bibfnamefont\endcsname\relax
  \def\bibfnamefont#1{#1}\fi
\expandafter\ifx\csname citenamefont\endcsname\relax
  \def\citenamefont#1{#1}\fi
\expandafter\ifx\csname url\endcsname\relax
  \def\url#1{\texttt{#1}}\fi
\expandafter\ifx\csname urlprefix\endcsname\relax\def\urlprefix{URL }\fi
\providecommand{\bibinfo}[2]{#2}
\providecommand{\eprint}[2][]{\url{#2}}

\bibitem[{\citenamefont{Van~Hove}(1954)}]{VanHove1954}
\bibinfo{author}{\bibfnamefont{L.}~\bibnamefont{Van~Hove}},
  \bibinfo{journal}{Phys. Rev.} \textbf{\bibinfo{volume}{95}},
  \bibinfo{pages}{249} (\bibinfo{year}{1954}).

\bibitem[{\citenamefont{Lovesey}(2003)}]{Lovesey2003}
\bibinfo{author}{\bibfnamefont{S.~W.} \bibnamefont{Lovesey}},
  \emph{\bibinfo{title}{Theory of Neutron Scattering from Condensed Matter}}
  (\bibinfo{publisher}{Oxford}, \bibinfo{year}{2003}).

\bibitem[{\citenamefont{Schermer and Blume}(1968)}]{Schermer1968}
\bibinfo{author}{\bibfnamefont{R.~I.} \bibnamefont{Schermer}} \bibnamefont{and}
  \bibinfo{author}{\bibfnamefont{M.}~\bibnamefont{Blume}},
  \bibinfo{journal}{Phys. Rev.} \textbf{\bibinfo{volume}{166}},
  \bibinfo{pages}{554} (\bibinfo{year}{1968}).

\bibitem[{\citenamefont{Walker and Happer}(1997)}]{WalkerHapper}
\bibinfo{author}{\bibfnamefont{T.~G.} \bibnamefont{Walker}} \bibnamefont{and}
  \bibinfo{author}{\bibfnamefont{W.}~\bibnamefont{Happer}},
  \bibinfo{journal}{Rev. Mod. Phys.} \textbf{\bibinfo{volume}{69}},
  \bibinfo{pages}{629} (\bibinfo{year}{1997}).

\bibitem[{\citenamefont{Driehuys et~al.}(1996)\citenamefont{Driehuys, Cates,
  Miron, Sauer, Walter, and Happer}}]{bascontflow}
\bibinfo{author}{\bibfnamefont{B.}~\bibnamefont{Driehuys}},
  \bibinfo{author}{\bibfnamefont{G.~D.} \bibnamefont{Cates}},
  \bibinfo{author}{\bibfnamefont{E.}~\bibnamefont{Miron}},
  \bibinfo{author}{\bibfnamefont{K.}~\bibnamefont{Sauer}},
  \bibinfo{author}{\bibfnamefont{D.~K.} \bibnamefont{Walter}},
  \bibnamefont{and} \bibinfo{author}{\bibfnamefont{W.}~\bibnamefont{Happer}},
  \bibinfo{journal}{Appl. Phys. Lett.} \textbf{\bibinfo{volume}{69}},
  \bibinfo{pages}{1668} (\bibinfo{year}{1996}).

\bibitem[{\citenamefont{Rosen et~al.}(1999)\citenamefont{Rosen, Chupp, Coulter,
  Welsh, and Swanson}}]{rosen1999}
\bibinfo{author}{\bibfnamefont{M.~S.} \bibnamefont{Rosen}},
  \bibinfo{author}{\bibfnamefont{T.~E.} \bibnamefont{Chupp}},
  \bibinfo{author}{\bibfnamefont{K.~P.} \bibnamefont{Coulter}},
  \bibinfo{author}{\bibfnamefont{R.~C.} \bibnamefont{Welsh}}, \bibnamefont{and}
  \bibinfo{author}{\bibfnamefont{S.~D.} \bibnamefont{Swanson}},
  \bibinfo{journal}{Rev. Sci. Inst.} \textbf{\bibinfo{volume}{70}},
  \bibinfo{pages}{1546} (\bibinfo{year}{1999}).

\bibitem[{\citenamefont{Zook et~al.}(2002)\citenamefont{Zook, Adhyaru, and
  Bowers}}]{zook}
\bibinfo{author}{\bibfnamefont{A.~L.} \bibnamefont{Zook}},
  \bibinfo{author}{\bibfnamefont{B.~B.} \bibnamefont{Adhyaru}},
  \bibnamefont{and} \bibinfo{author}{\bibfnamefont{C.~R.}
  \bibnamefont{Bowers}}, \bibinfo{journal}{J. Magn. Reson.}
  \textbf{\bibinfo{volume}{159}}, \bibinfo{pages}{175} (\bibinfo{year}{2002}).

\bibitem[{\citenamefont{Mortuza et~al.}(2003)\citenamefont{Mortuza, Anala,
  Pavlovskaya, Dieken, and Meersmann}}]{meersmann129}
\bibinfo{author}{\bibfnamefont{M.~G.} \bibnamefont{Mortuza}},
  \bibinfo{author}{\bibfnamefont{S.}~\bibnamefont{Anala}},
  \bibinfo{author}{\bibfnamefont{G.~E.} \bibnamefont{Pavlovskaya}},
  \bibinfo{author}{\bibfnamefont{T.~J.} \bibnamefont{Dieken}},
  \bibnamefont{and}
  \bibinfo{author}{\bibfnamefont{T.}~\bibnamefont{Meersmann}},
  \bibinfo{journal}{J. Chem. Phys.} \textbf{\bibinfo{volume}{118}},
  \bibinfo{pages}{1581} (\bibinfo{year}{2003}).

\bibitem[{\citenamefont{Ruset et~al.}(2006)\citenamefont{Ruset, Ketel, and
  Hersman}}]{hersmanPRL}
\bibinfo{author}{\bibfnamefont{I.~C.} \bibnamefont{Ruset}},
  \bibinfo{author}{\bibfnamefont{S.}~\bibnamefont{Ketel}}, \bibnamefont{and}
  \bibinfo{author}{\bibfnamefont{F.~W.} \bibnamefont{Hersman}},
  \bibinfo{journal}{Phys. Rev. Lett.} \textbf{\bibinfo{volume}{96}},
  \bibinfo{pages}{053002} (\bibinfo{year}{2006}).

\bibitem[{\citenamefont{Nikolaou et~al.}(2013)\citenamefont{Nikolaou, Coffey,
  Walkup, Gust, Whiting, Newton, Barcus, Muradyan, Dabaghyan, Moroz
  et~al.}}]{NikolaouhighP2}
\bibinfo{author}{\bibfnamefont{P.}~\bibnamefont{Nikolaou}},
  \bibinfo{author}{\bibfnamefont{A.~M.} \bibnamefont{Coffey}},
  \bibinfo{author}{\bibfnamefont{L.~L.} \bibnamefont{Walkup}},
  \bibinfo{author}{\bibfnamefont{B.~M.} \bibnamefont{Gust}},
  \bibinfo{author}{\bibfnamefont{N.}~\bibnamefont{Whiting}},
  \bibinfo{author}{\bibfnamefont{H.}~\bibnamefont{Newton}},
  \bibinfo{author}{\bibfnamefont{S.}~\bibnamefont{Barcus}},
  \bibinfo{author}{\bibfnamefont{I.}~\bibnamefont{Muradyan}},
  \bibinfo{author}{\bibfnamefont{M.}~\bibnamefont{Dabaghyan}},
  \bibinfo{author}{\bibfnamefont{G.~D.} \bibnamefont{Moroz}},
  \bibnamefont{et~al.}, \bibinfo{journal}{Proc. Natl. Acad. Sci. USA}
  \textbf{\bibinfo{volume}{86}}, \bibinfo{pages}{14150} (\bibinfo{year}{2013}).

\bibitem[{\citenamefont{Nikolaou et~al.}(2014)\citenamefont{Nikolaou, Coffey,
  Barlow, Rosen, Goodson, and Chekmenev}}]{NikolaouhighP}
\bibinfo{author}{\bibfnamefont{P.}~\bibnamefont{Nikolaou}},
  \bibinfo{author}{\bibfnamefont{A.~M.} \bibnamefont{Coffey}},
  \bibinfo{author}{\bibfnamefont{M.~J.} \bibnamefont{Barlow}},
  \bibinfo{author}{\bibfnamefont{M.~S.} \bibnamefont{Rosen}},
  \bibinfo{author}{\bibfnamefont{B.~M.} \bibnamefont{Goodson}},
  \bibnamefont{and} \bibinfo{author}{\bibfnamefont{E.~Y.}
  \bibnamefont{Chekmenev}}, \bibinfo{journal}{Anal. Chem.}
  \textbf{\bibinfo{volume}{86}}, \bibinfo{pages}{8206} (\bibinfo{year}{2014}).

\bibitem[{\citenamefont{Norquay et~al.}(2018)\citenamefont{Norquay, Collier,
  Rao, Stewart, and Wild}}]{sheffield}
\bibinfo{author}{\bibfnamefont{G.}~\bibnamefont{Norquay}},
  \bibinfo{author}{\bibfnamefont{G.~J.} \bibnamefont{Collier}},
  \bibinfo{author}{\bibfnamefont{M.}~\bibnamefont{Rao}},
  \bibinfo{author}{\bibfnamefont{N.~J.} \bibnamefont{Stewart}},
  \bibnamefont{and} \bibinfo{author}{\bibfnamefont{J.~M.} \bibnamefont{Wild}},
  \bibinfo{journal}{Phys. Rev. Lett.} \textbf{\bibinfo{volume}{121}},
  \bibinfo{pages}{153201} (\bibinfo{year}{2018}).

\bibitem[{\citenamefont{Gatzke et~al.}(1993)\citenamefont{Gatzke, Cates,
  Driehuys, Fox, Happer, and Saam}}]{gatzke1993}
\bibinfo{author}{\bibfnamefont{M.}~\bibnamefont{Gatzke}},
  \bibinfo{author}{\bibfnamefont{G.~D.} \bibnamefont{Cates}},
  \bibinfo{author}{\bibfnamefont{B.}~\bibnamefont{Driehuys}},
  \bibinfo{author}{\bibfnamefont{D.}~\bibnamefont{Fox}},
  \bibinfo{author}{\bibfnamefont{W.}~\bibnamefont{Happer}}, \bibnamefont{and}
  \bibinfo{author}{\bibfnamefont{B.}~\bibnamefont{Saam}},
  \bibinfo{journal}{Phys. Rev. Lett.} \textbf{\bibinfo{volume}{70}},
  \bibinfo{pages}{690} (\bibinfo{year}{1993}).

\bibitem[{\citenamefont{Sauer et~al.}(1997)\citenamefont{Sauer, Fitzgerald, and
  Happer}}]{Sauer1997}
\bibinfo{author}{\bibfnamefont{K.}~\bibnamefont{Sauer}},
  \bibinfo{author}{\bibfnamefont{R.}~\bibnamefont{Fitzgerald}},
  \bibnamefont{and} \bibinfo{author}{\bibfnamefont{W.}~\bibnamefont{Happer}},
  \bibinfo{journal}{Chem. Phys. Lett.} \textbf{\bibinfo{volume}{277}},
  \bibinfo{pages}{153} (\bibinfo{year}{1997}).

\bibitem[{\citenamefont{Haake et~al.}(1998)\citenamefont{Haake, Goodson, Laws,
  Brunner, Cyrier, Havlin, and Pines}}]{Haake1998}
\bibinfo{author}{\bibfnamefont{M.}~\bibnamefont{Haake}},
  \bibinfo{author}{\bibfnamefont{B.}~\bibnamefont{Goodson}},
  \bibinfo{author}{\bibfnamefont{D.}~\bibnamefont{Laws}},
  \bibinfo{author}{\bibfnamefont{E.}~\bibnamefont{Brunner}},
  \bibinfo{author}{\bibfnamefont{M.}~\bibnamefont{Cyrier}},
  \bibinfo{author}{\bibfnamefont{R.}~\bibnamefont{Havlin}}, \bibnamefont{and}
  \bibinfo{author}{\bibfnamefont{A.}~\bibnamefont{Pines}},
  \bibinfo{journal}{Chem. Phys. Lett.} \textbf{\bibinfo{volume}{292}},
  \bibinfo{pages}{686} (\bibinfo{year}{1998}).

\bibitem[{\citenamefont{Katz et~al.}(2022{\natexlab{a}})\citenamefont{Katz,
  Shaham, and Firstenberg}}]{Katz2022a}
\bibinfo{author}{\bibfnamefont{O.}~\bibnamefont{Katz}},
  \bibinfo{author}{\bibfnamefont{R.}~\bibnamefont{Shaham}}, \bibnamefont{and}
  \bibinfo{author}{\bibfnamefont{O.}~\bibnamefont{Firstenberg}},
  \bibinfo{journal}{PRX Quantum} \textbf{\bibinfo{volume}{3}},
  \bibinfo{pages}{010305} (\bibinfo{year}{2022}{\natexlab{a}}).

\bibitem[{\citenamefont{Katz et~al.}(2022{\natexlab{b}})\citenamefont{Katz,
  Shaham, Reches, Gorshkov, and Firstenberg}}]{Katz2022b}
\bibinfo{author}{\bibfnamefont{O.}~\bibnamefont{Katz}},
  \bibinfo{author}{\bibfnamefont{R.}~\bibnamefont{Shaham}},
  \bibinfo{author}{\bibfnamefont{E.}~\bibnamefont{Reches}},
  \bibinfo{author}{\bibfnamefont{A.~V.} \bibnamefont{Gorshkov}},
  \bibnamefont{and}
  \bibinfo{author}{\bibfnamefont{O.}~\bibnamefont{Firstenberg}},
  \bibinfo{journal}{Phys. Rev. A} \textbf{\bibinfo{volume}{105}},
  \bibinfo{pages}{042606} (\bibinfo{year}{2022}{\natexlab{b}}).

\bibitem[{\citenamefont{Goodson}(2002)}]{boydrev2002}
\bibinfo{author}{\bibfnamefont{B.~M.} \bibnamefont{Goodson}},
  \bibinfo{journal}{J. Magn. Reson.} \textbf{\bibinfo{volume}{155}},
  \bibinfo{pages}{157} (\bibinfo{year}{2002}).

\bibitem[{\citenamefont{Mugler~III and Altes}(2013)}]{lungrev}
\bibinfo{author}{\bibfnamefont{J.~P.} \bibnamefont{Mugler~III}}
  \bibnamefont{and} \bibinfo{author}{\bibfnamefont{T.~A.} \bibnamefont{Altes}},
  \bibinfo{journal}{J. Magn. Reson. Imag.} \textbf{\bibinfo{volume}{37}},
  \bibinfo{pages}{313} (\bibinfo{year}{2013}).

\bibitem[{\citenamefont{Khan et~al.}(2021)\citenamefont{Khan, Harvey, Birchall,
  Irwin, Nikolaou, Schrank, Emami, Dummer, Barlow, Goodson et~al.}}]{Khan2021}
\bibinfo{author}{\bibfnamefont{A.}~\bibnamefont{Khan}},
  \bibinfo{author}{\bibfnamefont{R.}~\bibnamefont{Harvey}},
  \bibinfo{author}{\bibfnamefont{J.}~\bibnamefont{Birchall}},
  \bibinfo{author}{\bibfnamefont{R.}~\bibnamefont{Irwin}},
  \bibinfo{author}{\bibfnamefont{P.}~\bibnamefont{Nikolaou}},
  \bibinfo{author}{\bibfnamefont{G.}~\bibnamefont{Schrank}},
  \bibinfo{author}{\bibfnamefont{K.}~\bibnamefont{Emami}},
  \bibinfo{author}{\bibfnamefont{A.}~\bibnamefont{Dummer}},
  \bibinfo{author}{\bibfnamefont{M.}~\bibnamefont{Barlow}},
  \bibinfo{author}{\bibfnamefont{B.}~\bibnamefont{Goodson}},
  \bibnamefont{et~al.}, \bibinfo{journal}{Angew. Chem. Int. Ed.}
  \textbf{\bibinfo{volume}{60}}, \bibinfo{pages}{22126} (\bibinfo{year}{2021}).

\bibitem[{\citenamefont{Zheng et~al.}(2016)\citenamefont{Zheng, Miller, Tobias,
  and Cates}}]{meta131nature}
\bibinfo{author}{\bibfnamefont{Y.}~\bibnamefont{Zheng}},
  \bibinfo{author}{\bibfnamefont{G.~W.} \bibnamefont{Miller}},
  \bibinfo{author}{\bibfnamefont{W.~A.} \bibnamefont{Tobias}},
  \bibnamefont{and} \bibinfo{author}{\bibfnamefont{G.~D.} \bibnamefont{Cates}},
  \bibinfo{journal}{Nature} \textbf{\bibinfo{volume}{537}},
  \bibinfo{pages}{652} (\bibinfo{year}{2016}).

\bibitem[{\citenamefont{Barskiy et~al.}(2017)\citenamefont{Barskiy, Coffey,
  Nikolaou, Mikhaylov, Goodson, Branca, Lu, Shapiro, Telkki, Zhivonitko
  et~al.}}]{danillarev}
\bibinfo{author}{\bibfnamefont{D.~A.} \bibnamefont{Barskiy}},
  \bibinfo{author}{\bibfnamefont{A.~M.} \bibnamefont{Coffey}},
  \bibinfo{author}{\bibfnamefont{P.}~\bibnamefont{Nikolaou}},
  \bibinfo{author}{\bibfnamefont{D.~M.} \bibnamefont{Mikhaylov}},
  \bibinfo{author}{\bibfnamefont{B.~M.} \bibnamefont{Goodson}},
  \bibinfo{author}{\bibfnamefont{R.~T.} \bibnamefont{Branca}},
  \bibinfo{author}{\bibfnamefont{G.~J.} \bibnamefont{Lu}},
  \bibinfo{author}{\bibfnamefont{M.~G.} \bibnamefont{Shapiro}},
  \bibinfo{author}{\bibfnamefont{V.-V.} \bibnamefont{Telkki}},
  \bibinfo{author}{\bibfnamefont{V.~V.} \bibnamefont{Zhivonitko}},
  \bibnamefont{et~al.}, \bibinfo{journal}{Chem. Eur. J.}
  \textbf{\bibinfo{volume}{23}}, \bibinfo{pages}{725} (\bibinfo{year}{2017}).

\bibitem[{\citenamefont{Wu et~al.}(1987)\citenamefont{Wu, Happer, and
  Daniels}}]{Wu1987}
\bibinfo{author}{\bibfnamefont{Z.}~\bibnamefont{Wu}},
  \bibinfo{author}{\bibfnamefont{W.}~\bibnamefont{Happer}}, \bibnamefont{and}
  \bibinfo{author}{\bibfnamefont{J.~M.} \bibnamefont{Daniels}},
  \bibinfo{journal}{Phys. Rev. Lett.} \textbf{\bibinfo{volume}{59}},
  \bibinfo{pages}{1480} (\bibinfo{year}{1987}).

\bibitem[{\citenamefont{Wu et~al.}(1988)\citenamefont{Wu, Schaefer, Cates, and
  Happer}}]{Wu1988}
\bibinfo{author}{\bibfnamefont{Z.}~\bibnamefont{Wu}},
  \bibinfo{author}{\bibfnamefont{S.}~\bibnamefont{Schaefer}},
  \bibinfo{author}{\bibfnamefont{G.}~\bibnamefont{Cates}}, \bibnamefont{and}
  \bibinfo{author}{\bibfnamefont{W.}~\bibnamefont{Happer}},
  \bibinfo{journal}{Phys. Rev. A} \textbf{\bibinfo{volume}{37}},
  \bibinfo{pages}{1161} (\bibinfo{year}{1988}).

\bibitem[{\citenamefont{Raftery et~al.}(1991)\citenamefont{Raftery, Long,
  Meersmann, Grandinetti, Reven, and Pines}}]{Raftery1991}
\bibinfo{author}{\bibfnamefont{D.}~\bibnamefont{Raftery}},
  \bibinfo{author}{\bibfnamefont{H.}~\bibnamefont{Long}},
  \bibinfo{author}{\bibfnamefont{T.}~\bibnamefont{Meersmann}},
  \bibinfo{author}{\bibfnamefont{P.~J.} \bibnamefont{Grandinetti}},
  \bibinfo{author}{\bibfnamefont{L.}~\bibnamefont{Reven}}, \bibnamefont{and}
  \bibinfo{author}{\bibfnamefont{A.}~\bibnamefont{Pines}},
  \bibinfo{journal}{Phys. Rev. Lett.} \textbf{\bibinfo{volume}{66}},
  \bibinfo{pages}{584} (\bibinfo{year}{1991}).

\bibitem[{\citenamefont{Appelt et~al.}(1994)\citenamefont{Appelt, W{\"a}ckerle,
  and Mehring}}]{mehringgyro}
\bibinfo{author}{\bibfnamefont{S.}~\bibnamefont{Appelt}},
  \bibinfo{author}{\bibfnamefont{G.}~\bibnamefont{W{\"a}ckerle}},
  \bibnamefont{and} \bibinfo{author}{\bibfnamefont{M.}~\bibnamefont{Mehring}},
  \bibinfo{journal}{Phys. Rev. Lett.} \textbf{\bibinfo{volume}{72}},
  \bibinfo{pages}{3921} (\bibinfo{year}{1994}).

\bibitem[{\citenamefont{Bear et~al.}(2000)\citenamefont{Bear, Stoner,
  Walsworth, Kostelecky, and Lane}}]{bear2000}
\bibinfo{author}{\bibfnamefont{D.}~\bibnamefont{Bear}},
  \bibinfo{author}{\bibfnamefont{R.~E.} \bibnamefont{Stoner}},
  \bibinfo{author}{\bibfnamefont{R.~L.} \bibnamefont{Walsworth}},
  \bibinfo{author}{\bibfnamefont{V.~A.} \bibnamefont{Kostelecky}},
  \bibnamefont{and} \bibinfo{author}{\bibfnamefont{C.~D.} \bibnamefont{Lane}},
  \bibinfo{journal}{Phys. Rev. Lett.} \textbf{\bibinfo{volume}{85}},
  \bibinfo{pages}{5038} (\bibinfo{year}{2000}).

\bibitem[{\citenamefont{Cane et~al.}(2004)\citenamefont{Cane, Bear, Phillips,
  Rosen, Smallwood, Stoner, Walsworth, and Kostelecky}}]{cane2004}
\bibinfo{author}{\bibfnamefont{F.}~\bibnamefont{Cane}},
  \bibinfo{author}{\bibfnamefont{D.}~\bibnamefont{Bear}},
  \bibinfo{author}{\bibfnamefont{D.~F.} \bibnamefont{Phillips}},
  \bibinfo{author}{\bibfnamefont{M.~S.} \bibnamefont{Rosen}},
  \bibinfo{author}{\bibfnamefont{C.~L.} \bibnamefont{Smallwood}},
  \bibinfo{author}{\bibfnamefont{R.~E.} \bibnamefont{Stoner}},
  \bibinfo{author}{\bibfnamefont{R.~L.} \bibnamefont{Walsworth}},
  \bibnamefont{and} \bibinfo{author}{\bibfnamefont{V.~A.}
  \bibnamefont{Kostelecky}}, \bibinfo{journal}{Phys. Rev. Lett.}
  \textbf{\bibinfo{volume}{93}}, \bibinfo{pages}{230801}
  (\bibinfo{year}{2004}).

\bibitem[{\citenamefont{Gemmel et~al.}(2010)\citenamefont{Gemmel, Heil, Karpuk,
  Lenz, Sobolev, Tullney, Burghoff, Kilian, Knappe-Gruneberg, Muller
  et~al.}}]{gemmel2010}
\bibinfo{author}{\bibfnamefont{C.}~\bibnamefont{Gemmel}},
  \bibinfo{author}{\bibfnamefont{W.}~\bibnamefont{Heil}},
  \bibinfo{author}{\bibfnamefont{S.}~\bibnamefont{Karpuk}},
  \bibinfo{author}{\bibfnamefont{K.}~\bibnamefont{Lenz}},
  \bibinfo{author}{\bibfnamefont{Y.}~\bibnamefont{Sobolev}},
  \bibinfo{author}{\bibfnamefont{K.}~\bibnamefont{Tullney}},
  \bibinfo{author}{\bibfnamefont{M.}~\bibnamefont{Burghoff}},
  \bibinfo{author}{\bibfnamefont{W.}~\bibnamefont{Kilian}},
  \bibinfo{author}{\bibfnamefont{S.}~\bibnamefont{Knappe-Gruneberg}},
  \bibinfo{author}{\bibfnamefont{W.}~\bibnamefont{Muller}},
  \bibnamefont{et~al.}, \bibinfo{journal}{Phys. Rev. D}
  \textbf{\bibinfo{volume}{82}}, \bibinfo{pages}{111901(R)}
  (\bibinfo{year}{2010}).

\bibitem[{\citenamefont{Allmendinger et~al.}(2014)\citenamefont{Allmendinger,
  Heil, Karpuk, Kilian, Scharth, Schmidt, Schnabel, Sobolev, and
  Tullney}}]{Allmendinger2014}
\bibinfo{author}{\bibfnamefont{F.}~\bibnamefont{Allmendinger}},
  \bibinfo{author}{\bibfnamefont{W.}~\bibnamefont{Heil}},
  \bibinfo{author}{\bibfnamefont{S.}~\bibnamefont{Karpuk}},
  \bibinfo{author}{\bibfnamefont{W.}~\bibnamefont{Kilian}},
  \bibinfo{author}{\bibfnamefont{A.}~\bibnamefont{Scharth}},
  \bibinfo{author}{\bibfnamefont{U.}~\bibnamefont{Schmidt}},
  \bibinfo{author}{\bibfnamefont{A.}~\bibnamefont{Schnabel}},
  \bibinfo{author}{\bibfnamefont{Y.}~\bibnamefont{Sobolev}}, \bibnamefont{and}
  \bibinfo{author}{\bibfnamefont{K.}~\bibnamefont{Tullney}},
  \bibinfo{journal}{Phys. Rev. Lett.} \textbf{\bibinfo{volume}{112}},
  \bibinfo{pages}{110801} (\bibinfo{year}{2014}).

\bibitem[{\citenamefont{Stadnik and Flambaum}(2015)}]{Stadnik2015}
\bibinfo{author}{\bibfnamefont{Y.~V.} \bibnamefont{Stadnik}} \bibnamefont{and}
  \bibinfo{author}{\bibfnamefont{V.~V.} \bibnamefont{Flambaum}},
  \bibinfo{journal}{Eur. Phys. J. C} \textbf{\bibinfo{volume}{75}},
  \bibinfo{pages}{110} (\bibinfo{year}{2015}).

\bibitem[{\citenamefont{Kostelecky and Vargas}(2018)}]{Kostelecky2018}
\bibinfo{author}{\bibfnamefont{V.~A.} \bibnamefont{Kostelecky}}
  \bibnamefont{and} \bibinfo{author}{\bibfnamefont{A.~J.}
  \bibnamefont{Vargas}}, \bibinfo{journal}{Phys. Rev. D}
  \textbf{\bibinfo{volume}{98}}, \bibinfo{pages}{036003}
  (\bibinfo{year}{2018}).

\bibitem[{\citenamefont{Sachdeva et~al.}(2019)\citenamefont{Sachdeva, Fan,
  Babcock, Burghoff, Chupp, Degenkolb, Fierlinger, Haude, Kraegeloh, Kilian
  et~al.}}]{Sachdeva2019}
\bibinfo{author}{\bibfnamefont{N.}~\bibnamefont{Sachdeva}},
  \bibinfo{author}{\bibfnamefont{I.}~\bibnamefont{Fan}},
  \bibinfo{author}{\bibfnamefont{E.}~\bibnamefont{Babcock}},
  \bibinfo{author}{\bibfnamefont{M.}~\bibnamefont{Burghoff}},
  \bibinfo{author}{\bibfnamefont{T.~E.} \bibnamefont{Chupp}},
  \bibinfo{author}{\bibfnamefont{S.}~\bibnamefont{Degenkolb}},
  \bibinfo{author}{\bibfnamefont{P.}~\bibnamefont{Fierlinger}},
  \bibinfo{author}{\bibfnamefont{S.}~\bibnamefont{Haude}},
  \bibinfo{author}{\bibfnamefont{E.}~\bibnamefont{Kraegeloh}},
  \bibinfo{author}{\bibfnamefont{W.}~\bibnamefont{Kilian}},
  \bibnamefont{et~al.}, \bibinfo{journal}{Phys. Rev. Lett.}
  \textbf{\bibinfo{volume}{123}}, \bibinfo{pages}{143003}
  (\bibinfo{year}{2019}).

\bibitem[{\citenamefont{Bulatowicz et~al.}(2013)\citenamefont{Bulatowicz,
  Griffith, Larsen, Mirijanian, Fu, Smith, Snow, Yan, and
  Walker}}]{walkerNsnow}
\bibinfo{author}{\bibfnamefont{M.}~\bibnamefont{Bulatowicz}},
  \bibinfo{author}{\bibfnamefont{R.}~\bibnamefont{Griffith}},
  \bibinfo{author}{\bibfnamefont{M.}~\bibnamefont{Larsen}},
  \bibinfo{author}{\bibfnamefont{J.}~\bibnamefont{Mirijanian}},
  \bibinfo{author}{\bibfnamefont{C.~B.} \bibnamefont{Fu}},
  \bibinfo{author}{\bibfnamefont{E.}~\bibnamefont{Smith}},
  \bibinfo{author}{\bibfnamefont{W.~M.} \bibnamefont{Snow}},
  \bibinfo{author}{\bibfnamefont{H.}~\bibnamefont{Yan}}, \bibnamefont{and}
  \bibinfo{author}{\bibfnamefont{T.~G.} \bibnamefont{Walker}},
  \bibinfo{journal}{Phys. Rev. Lett.} \textbf{\bibinfo{volume}{111}},
  \bibinfo{pages}{102001} (\bibinfo{year}{2013}).

\bibitem[{\citenamefont{Feng et~al.}(2022)\citenamefont{Feng, Ning, Zhang, Lu,
  and Sheng}}]{Feng:2022}
\bibinfo{author}{\bibfnamefont{Y.-K.} \bibnamefont{Feng}},
  \bibinfo{author}{\bibfnamefont{D.-H.} \bibnamefont{Ning}},
  \bibinfo{author}{\bibfnamefont{S.-B.} \bibnamefont{Zhang}},
  \bibinfo{author}{\bibfnamefont{Z.-T.} \bibnamefont{Lu}}, \bibnamefont{and}
  \bibinfo{author}{\bibfnamefont{D.}~\bibnamefont{Sheng}},
  \bibinfo{journal}{Phys. Rev. Lett.} \textbf{\bibinfo{volume}{128}},
  \bibinfo{pages}{231803} (\bibinfo{year}{2022}),
  \urlprefix\url{https://link.aps.org/doi/10.1103/PhysRevLett.128.231803}.

\bibitem[{\citenamefont{Gudkov and Shimizu}(2020)}]{Gudkov20}
\bibinfo{author}{\bibfnamefont{V.}~\bibnamefont{Gudkov}} \bibnamefont{and}
  \bibinfo{author}{\bibfnamefont{H.~M.} \bibnamefont{Shimizu}},
  \bibinfo{journal}{Phys. Rev. C} \textbf{\bibinfo{volume}{102}},
  \bibinfo{pages}{015503} (\bibinfo{year}{2020}).

\bibitem[{\citenamefont{Sears}(1989)}]{Sears}
\bibinfo{author}{\bibfnamefont{V.~F.} \bibnamefont{Sears}},
  \emph{\bibinfo{title}{Neutron Optics: An Introduction to the Theory of
  Neutron Optical Phenomena and Their Applications}}
  (\bibinfo{publisher}{Oxford University Press}, \bibinfo{address}{New York},
  \bibinfo{year}{1989}).

\bibitem[{\citenamefont{Baryshevsky and Podgoretsky}(1964)}]{Baryshevsky1965}
\bibinfo{author}{\bibfnamefont{V.}~\bibnamefont{Baryshevsky}} \bibnamefont{and}
  \bibinfo{author}{\bibfnamefont{M.}~\bibnamefont{Podgoretsky}},
  \bibinfo{journal}{Zh. Eksp. Teor. Fiz.} \textbf{\bibinfo{volume}{47}},
  \bibinfo{pages}{1050} (\bibinfo{year}{1964}).

\bibitem[{\citenamefont{Abragam and Goldman}(1982)}]{Abragam1982}
\bibinfo{author}{\bibfnamefont{A.}~\bibnamefont{Abragam}} \bibnamefont{and}
  \bibinfo{author}{\bibfnamefont{M.}~\bibnamefont{Goldman}},
  \emph{\bibinfo{title}{Nuclear magnetism: order and disorder}}
  (\bibinfo{publisher}{Oxford}, \bibinfo{year}{1982}).

\bibitem[{\citenamefont{Ramsey}(1956)}]{Ramsey1956}
\bibinfo{author}{\bibfnamefont{N.~F.} \bibnamefont{Ramsey}},
  \emph{\bibinfo{title}{Molecular Beams}} (\bibinfo{publisher}{Oxford},
  \bibinfo{year}{1956}).

\bibitem[{\citenamefont{Ramsey}(1990)}]{Ramsey1990}
\bibinfo{author}{\bibfnamefont{N.~F.} \bibnamefont{Ramsey}},
  \bibinfo{journal}{Rev. Mod. Phys.} \textbf{\bibinfo{volume}{62}},
  \bibinfo{pages}{541} (\bibinfo{year}{1990}).

\bibitem[{\citenamefont{Abragam et~al.}(1973)\citenamefont{Abragam, Bacchella,
  Gl{\"a}tti, Meriel, Pinot, and Piesvaux}}]{Abragam1973}
\bibinfo{author}{\bibfnamefont{A.}~\bibnamefont{Abragam}},
  \bibinfo{author}{\bibfnamefont{G.}~\bibnamefont{Bacchella}},
  \bibinfo{author}{\bibfnamefont{H.}~\bibnamefont{Gl{\"a}tti}},
  \bibinfo{author}{\bibfnamefont{P.}~\bibnamefont{Meriel}},
  \bibinfo{author}{\bibfnamefont{M.}~\bibnamefont{Pinot}}, \bibnamefont{and}
  \bibinfo{author}{\bibfnamefont{J.}~\bibnamefont{Piesvaux}},
  \bibinfo{journal}{Phys. Rev. Lett.} \textbf{\bibinfo{volume}{31}},
  \bibinfo{pages}{776} (\bibinfo{year}{1973}).

\bibitem[{\citenamefont{Forte}(1973)}]{Forte1973}
\bibinfo{author}{\bibfnamefont{M.}~\bibnamefont{Forte}},
  \bibinfo{journal}{Nuovo Cimento A} \textbf{\bibinfo{volume}{18A}},
  \bibinfo{pages}{726} (\bibinfo{year}{1973}).

\bibitem[{\citenamefont{Pokazanev and Skrotskii}(1979)}]{Pokazanev1979}
\bibinfo{author}{\bibfnamefont{V.~G.} \bibnamefont{Pokazanev}}
  \bibnamefont{and} \bibinfo{author}{\bibfnamefont{G.~V.}
  \bibnamefont{Skrotskii}}, \bibinfo{journal}{Phys.-Usp.}
  \textbf{\bibinfo{volume}{22}}, \bibinfo{pages}{943} (\bibinfo{year}{1979}).

\bibitem[{\citenamefont{Tsulaia}(2014)}]{Tsulaia2014}
\bibinfo{author}{\bibfnamefont{M.~I.} \bibnamefont{Tsulaia}},
  \bibinfo{journal}{Phys. At. Nucl.} \textbf{\bibinfo{volume}{77}},
  \bibinfo{pages}{1321} (\bibinfo{year}{2014}).

\bibitem[{\citenamefont{Mezei}(1972)}]{Mezei1972}
\bibinfo{author}{\bibfnamefont{F.}~\bibnamefont{Mezei}},
  \bibinfo{journal}{Zeitschrift fur Physik A Hadrons and Nuclei}
  \textbf{\bibinfo{volume}{255}}, \bibinfo{pages}{146} (\bibinfo{year}{1972}).

\bibitem[{\citenamefont{Zimmer et~al.}(2002)\citenamefont{Zimmer, Ehlers,
  Farago, Humblot, Ketter, and Scherm}}]{Zimmer2002}
\bibinfo{author}{\bibfnamefont{O.}~\bibnamefont{Zimmer}},
  \bibinfo{author}{\bibfnamefont{G.}~\bibnamefont{Ehlers}},
  \bibinfo{author}{\bibfnamefont{B.}~\bibnamefont{Farago}},
  \bibinfo{author}{\bibfnamefont{H.}~\bibnamefont{Humblot}},
  \bibinfo{author}{\bibfnamefont{W.}~\bibnamefont{Ketter}}, \bibnamefont{and}
  \bibinfo{author}{\bibfnamefont{R.}~\bibnamefont{Scherm}},
  \bibinfo{journal}{EPJ direct} \textbf{\bibinfo{volume}{4}},
  \bibinfo{pages}{1} (\bibinfo{year}{2002}).

\bibitem[{\citenamefont{Hahn}(1950)}]{Hahn1950}
\bibinfo{author}{\bibfnamefont{E.~L.} \bibnamefont{Hahn}},
  \bibinfo{journal}{Phys. Rev.} \textbf{\bibinfo{volume}{80}},
  \bibinfo{pages}{580} (\bibinfo{year}{1950}).

\bibitem[{\citenamefont{Zentrum}(2015)}]{neutrons}
\bibinfo{author}{\bibfnamefont{H.~M.-L.} \bibnamefont{Zentrum}},
  \bibinfo{journal}{Journal of large-scale research facilities}
  \textbf{\bibinfo{volume}{A11}}, \bibinfo{pages}{34} (\bibinfo{year}{2015}).

\bibitem[{\citenamefont{Pasini et~al.}(2019)\citenamefont{Pasini, Holderer,
  Kozielewski, Richter, and Monkenbusch}}]{Pasini2019}
\bibinfo{author}{\bibfnamefont{S.}~\bibnamefont{Pasini}},
  \bibinfo{author}{\bibfnamefont{O.}~\bibnamefont{Holderer}},
  \bibinfo{author}{\bibfnamefont{T.}~\bibnamefont{Kozielewski}},
  \bibinfo{author}{\bibfnamefont{D.}~\bibnamefont{Richter}}, \bibnamefont{and}
  \bibinfo{author}{\bibfnamefont{M.}~\bibnamefont{Monkenbusch}},
  \bibinfo{journal}{Review of Scientific Instruments}
  \textbf{\bibinfo{volume}{90}}, \bibinfo{pages}{043107}
  (\bibinfo{year}{2019}).

\bibitem[{\citenamefont{Rich et~al.}(2001)\citenamefont{Rich, Fan, Gentile,
  Hussey, Jones, Neff, Snow, and Thompson}}]{Rich2001}
\bibinfo{author}{\bibfnamefont{D.}~\bibnamefont{Rich}},
  \bibinfo{author}{\bibfnamefont{S.}~\bibnamefont{Fan}},
  \bibinfo{author}{\bibfnamefont{T.}~\bibnamefont{Gentile}},
  \bibinfo{author}{\bibfnamefont{D.}~\bibnamefont{Hussey}},
  \bibinfo{author}{\bibfnamefont{G.~L.} \bibnamefont{Jones}},
  \bibinfo{author}{\bibfnamefont{B.}~\bibnamefont{Neff}},
  \bibinfo{author}{\bibfnamefont{W.~M.} \bibnamefont{Snow}}, \bibnamefont{and}
  \bibinfo{author}{\bibfnamefont{A.}~\bibnamefont{Thompson}},
  \bibinfo{journal}{Physica B-condensed Matter} \textbf{\bibinfo{volume}{305}},
  \bibinfo{pages}{203} (\bibinfo{year}{2001}).

\bibitem[{\citenamefont{Salhi et~al.}(2014)\citenamefont{Salhi, Babcock,
  Pistel, and Ioffe}}]{Salhi2014}
\bibinfo{author}{\bibfnamefont{Z.}~\bibnamefont{Salhi}},
  \bibinfo{author}{\bibfnamefont{E.}~\bibnamefont{Babcock}},
  \bibinfo{author}{\bibfnamefont{P.}~\bibnamefont{Pistel}}, \bibnamefont{and}
  \bibinfo{author}{\bibfnamefont{A.}~\bibnamefont{Ioffe}},
  \bibinfo{journal}{Journal of Physics Conference Series}
  \textbf{\bibinfo{volume}{528}}, \bibinfo{pages}{012015}
  (\bibinfo{year}{2014}).

\bibitem[{\citenamefont{Molway et~al.}(2022)\citenamefont{Molway,
  Bales-Shaffer, Ranta, Basler, Murphy, Kidd, Gafar, Porter, Albin, Goodson
  et~al.}}]{Boyd1}
\bibinfo{author}{\bibfnamefont{M.}~\bibnamefont{Molway}},
  \bibinfo{author}{\bibfnamefont{L.}~\bibnamefont{Bales-Shaffer}},
  \bibinfo{author}{\bibfnamefont{K.}~\bibnamefont{Ranta}},
  \bibinfo{author}{\bibfnamefont{D.}~\bibnamefont{Basler}},
  \bibinfo{author}{\bibfnamefont{M.}~\bibnamefont{Murphy}},
  \bibinfo{author}{\bibfnamefont{B.}~\bibnamefont{Kidd}},
  \bibinfo{author}{\bibfnamefont{A.}~\bibnamefont{Gafar}},
  \bibinfo{author}{\bibfnamefont{J.}~\bibnamefont{Porter}},
  \bibinfo{author}{\bibfnamefont{K.}~\bibnamefont{Albin}},
  \bibinfo{author}{\bibfnamefont{B.}~\bibnamefont{Goodson}},
  \bibnamefont{et~al.}, \bibinfo{journal}{arXiv:2105.03076}
  (\bibinfo{year}{2022}).

\bibitem[{\citenamefont{Gafar et~al.}(2022)\citenamefont{Gafar, Basler,
  Cocking, Prince, Alam, Siraj, Petrilla, Rosen, Chekmenev, Babcock
  et~al.}}]{Boyd}
\bibinfo{author}{\bibfnamefont{A.}~\bibnamefont{Gafar}},
  \bibinfo{author}{\bibfnamefont{D.}~\bibnamefont{Basler}},
  \bibinfo{author}{\bibfnamefont{D.}~\bibnamefont{Cocking}},
  \bibinfo{author}{\bibfnamefont{M.}~\bibnamefont{Prince}},
  \bibinfo{author}{\bibfnamefont{M.~S.} \bibnamefont{Alam}},
  \bibinfo{author}{\bibfnamefont{Z.}~\bibnamefont{Siraj}},
  \bibinfo{author}{\bibfnamefont{A.}~\bibnamefont{Petrilla}},
  \bibinfo{author}{\bibfnamefont{M.}~\bibnamefont{Rosen}},
  \bibinfo{author}{\bibfnamefont{E.}~\bibnamefont{Chekmenev}},
  \bibinfo{author}{\bibfnamefont{E.}~\bibnamefont{Babcock}},
  \bibnamefont{et~al.}, \bibinfo{journal}{Manuscript in preparation}
  (\bibinfo{year}{2022}).

\bibitem[{\citenamefont{Babcock et~al.}(2016)\citenamefont{Babcock, Salhi,
  Theisselmann, Starostin, Schmeissner, Feoktystov, Mattauch, Pistel,
  Radulescu, and Ioffe}}]{Babcock2016}
\bibinfo{author}{\bibfnamefont{E.}~\bibnamefont{Babcock}},
  \bibinfo{author}{\bibfnamefont{Z.}~\bibnamefont{Salhi}},
  \bibinfo{author}{\bibfnamefont{T.}~\bibnamefont{Theisselmann}},
  \bibinfo{author}{\bibfnamefont{D.}~\bibnamefont{Starostin}},
  \bibinfo{author}{\bibfnamefont{J.}~\bibnamefont{Schmeissner}},
  \bibinfo{author}{\bibfnamefont{A.}~\bibnamefont{Feoktystov}},
  \bibinfo{author}{\bibfnamefont{S.}~\bibnamefont{Mattauch}},
  \bibinfo{author}{\bibfnamefont{P.}~\bibnamefont{Pistel}},
  \bibinfo{author}{\bibfnamefont{A.}~\bibnamefont{Radulescu}},
  \bibnamefont{and} \bibinfo{author}{\bibfnamefont{A.}~\bibnamefont{Ioffe}},
  \bibinfo{journal}{Journal of Physics Conference Series}
  \textbf{\bibinfo{volume}{711}}, \bibinfo{pages}{012008}
  (\bibinfo{year}{2016}).

\bibitem[{\citenamefont{Gentile et~al.}(2017)\citenamefont{Gentile, Nacher,
  Saam, and Walker}}]{Gentile2017}
\bibinfo{author}{\bibfnamefont{T.~R.} \bibnamefont{Gentile}},
  \bibinfo{author}{\bibfnamefont{P.~J.} \bibnamefont{Nacher}},
  \bibinfo{author}{\bibfnamefont{B.}~\bibnamefont{Saam}}, \bibnamefont{and}
  \bibinfo{author}{\bibfnamefont{T.~G.} \bibnamefont{Walker}},
  \bibinfo{journal}{Rev. Mod. Phys.} \textbf{\bibinfo{volume}{89}},
  \bibinfo{pages}{045004} (\bibinfo{year}{2017}).

\bibitem[{\citenamefont{Huber et~al.}(2014)\citenamefont{Huber, Arif, Chen,
  Gentile, Hussey, Black, Pushin, Shahi, and Yang}}]{Huber2014}
\bibinfo{author}{\bibfnamefont{M.~G.} \bibnamefont{Huber}},
  \bibinfo{author}{\bibfnamefont{M.}~\bibnamefont{Arif}},
  \bibinfo{author}{\bibfnamefont{W.~C.} \bibnamefont{Chen}},
  \bibinfo{author}{\bibfnamefont{T.~R.} \bibnamefont{Gentile}},
  \bibinfo{author}{\bibfnamefont{D.~S.} \bibnamefont{Hussey}},
  \bibinfo{author}{\bibfnamefont{T.~C.} \bibnamefont{Black}},
  \bibinfo{author}{\bibfnamefont{D.~A.} \bibnamefont{Pushin}},
  \bibinfo{author}{\bibfnamefont{F.~E.} \bibnamefont{Shahi},
  \bibfnamefont{C.~B.band~Wietfeldt}}, \bibnamefont{and}
  \bibinfo{author}{\bibfnamefont{L.}~\bibnamefont{Yang}},
  \bibinfo{journal}{Phys. Rev. C} \textbf{\bibinfo{volume}{90}},
  \bibinfo{pages}{064004} (\bibinfo{year}{2014}).

\bibitem[{\citenamefont{Huber}()}]{HuberPC}
\bibinfo{author}{\bibfnamefont{M.}~\bibnamefont{Huber}},
  \bibinfo{howpublished}{private communication}, \bibinfo{note}{full derivation
  of estimate of $\sigma_1$}.

\bibitem[{\citenamefont{Chupp et~al.}(2007)\citenamefont{Chupp, Coulter,
  Kandes, Sharma, Smith, Jones, Chen, Gentile, Rich, Lauss et~al.}}]{Chupp2007}
\bibinfo{author}{\bibfnamefont{T.}~\bibnamefont{Chupp}},
  \bibinfo{author}{\bibfnamefont{K.}~\bibnamefont{Coulter}},
  \bibinfo{author}{\bibfnamefont{M.}~\bibnamefont{Kandes}},
  \bibinfo{author}{\bibfnamefont{M.}~\bibnamefont{Sharma}},
  \bibinfo{author}{\bibfnamefont{T.}~\bibnamefont{Smith}},
  \bibinfo{author}{\bibfnamefont{G.~L.} \bibnamefont{Jones}},
  \bibinfo{author}{\bibfnamefont{W.}~\bibnamefont{Chen}},
  \bibinfo{author}{\bibfnamefont{T.}~\bibnamefont{Gentile}},
  \bibinfo{author}{\bibfnamefont{D.}~\bibnamefont{Rich}},
  \bibinfo{author}{\bibfnamefont{B.}~\bibnamefont{Lauss}},
  \bibnamefont{et~al.}, \bibinfo{journal}{Nucl. Instrumen. Meth. A}
  \textbf{\bibinfo{volume}{574}}, \bibinfo{pages}{500} (\bibinfo{year}{2007}).

\bibitem[{\citenamefont{Campbell et~al.}(2011)\citenamefont{Campbell, Wacklin,
  Sutton, Cubitt, and Fragneto}}]{Campbell2011}
\bibinfo{author}{\bibfnamefont{R.~A.} \bibnamefont{Campbell}},
  \bibinfo{author}{\bibfnamefont{H.~P.} \bibnamefont{Wacklin}},
  \bibinfo{author}{\bibfnamefont{I.}~\bibnamefont{Sutton}},
  \bibinfo{author}{\bibfnamefont{R.}~\bibnamefont{Cubitt}}, \bibnamefont{and}
  \bibinfo{author}{\bibfnamefont{G.}~\bibnamefont{Fragneto}},
  \bibinfo{journal}{European Physical Journal Plus}
  \textbf{\bibinfo{volume}{126}}, \bibinfo{pages}{107} (\bibinfo{year}{2011}).

\bibitem[{\citenamefont{Stupic et~al.}(2011)\citenamefont{Stupic, Cleveland,
  Pavlovskaya, and Meersmann}}]{Stupic2011}
\bibinfo{author}{\bibfnamefont{K.~F.} \bibnamefont{Stupic}},
  \bibinfo{author}{\bibfnamefont{Z.~I.} \bibnamefont{Cleveland}},
  \bibinfo{author}{\bibfnamefont{G.~E.} \bibnamefont{Pavlovskaya}},
  \bibnamefont{and}
  \bibinfo{author}{\bibfnamefont{T.}~\bibnamefont{Meersmann}},
  \bibinfo{journal}{J. Magn. Reson.} \textbf{\bibinfo{volume}{208}},
  \bibinfo{pages}{58} (\bibinfo{year}{2011}).

\bibitem[{\citenamefont{Lu et~al.}()\citenamefont{Lu, Snow, Goodson, Babcock,
  Salhi, Ioffe, Holderer, Passini, and Pistel}}]{HaoIP}
\bibinfo{author}{\bibfnamefont{H.}~\bibnamefont{Lu}},
  \bibinfo{author}{\bibfnamefont{W.~M.} \bibnamefont{Snow}},
  \bibinfo{author}{\bibfnamefont{B.~M.} \bibnamefont{Goodson}},
  \bibinfo{author}{\bibfnamefont{E.}~\bibnamefont{Babcock}},
  \bibinfo{author}{\bibfnamefont{Z.}~\bibnamefont{Salhi}},
  \bibinfo{author}{\bibfnamefont{A.}~\bibnamefont{Ioffe}},
  \bibinfo{author}{\bibfnamefont{O.}~\bibnamefont{Holderer}},
  \bibinfo{author}{\bibfnamefont{S.}~\bibnamefont{Passini}}, \bibnamefont{and}
  \bibinfo{author}{\bibfnamefont{P.}~\bibnamefont{Pistel}},
  \bibinfo{howpublished}{unpublished}, \bibinfo{note}{in process for PRC
  letters}.

\bibitem[{\citenamefont{Irfan et~al.}(2021)\citenamefont{Irfan, Backstone,
  Pynn, and Ortiz}}]{Irfan2021}
\bibinfo{author}{\bibfnamefont{A.~A.~M.} \bibnamefont{Irfan}},
  \bibinfo{author}{\bibfnamefont{P.}~\bibnamefont{Backstone}},
  \bibinfo{author}{\bibfnamefont{R.}~\bibnamefont{Pynn}}, \bibnamefont{and}
  \bibinfo{author}{\bibfnamefont{G.}~\bibnamefont{Ortiz}},
  \bibinfo{journal}{New J. Phys.} \textbf{\bibinfo{volume}{23}},
  \bibinfo{pages}{083022} (\bibinfo{year}{2021}).

\bibitem[{\citenamefont{Dantan et~al.}(2005)\citenamefont{Dantan, Reinaudi,
  Sinatra, Laloe, Giacobino, and Pinard}}]{Dantan2005}
\bibinfo{author}{\bibfnamefont{A.}~\bibnamefont{Dantan}},
  \bibinfo{author}{\bibfnamefont{G.}~\bibnamefont{Reinaudi}},
  \bibinfo{author}{\bibfnamefont{A.}~\bibnamefont{Sinatra}},
  \bibinfo{author}{\bibfnamefont{F.}~\bibnamefont{Laloe}},
  \bibinfo{author}{\bibfnamefont{E.}~\bibnamefont{Giacobino}},
  \bibnamefont{and} \bibinfo{author}{\bibfnamefont{M.}~\bibnamefont{Pinard}},
  \bibinfo{journal}{Phys. Rev. Lett.} \textbf{\bibinfo{volume}{95}},
  \bibinfo{pages}{123002} (\bibinfo{year}{2005}).

\bibitem[{\citenamefont{Katz et~al.}(2021)\citenamefont{Katz, Shaham, and
  Firstenberg}}]{Katz2021}
\bibinfo{author}{\bibfnamefont{O.}~\bibnamefont{Katz}},
  \bibinfo{author}{\bibfnamefont{R.}~\bibnamefont{Shaham}}, \bibnamefont{and}
  \bibinfo{author}{\bibfnamefont{O.}~\bibnamefont{Firstenberg}},
  \bibinfo{journal}{Sci. Adv.} \textbf{\bibinfo{volume}{7}},
  \bibinfo{pages}{eabe9164} (\bibinfo{year}{2021}).

\bibitem[{\citenamefont{Katz et~al.}(2020)\citenamefont{Katz, Shaham, Polzik,
  and Firstenberg}}]{Katz2020}
\bibinfo{author}{\bibfnamefont{O.}~\bibnamefont{Katz}},
  \bibinfo{author}{\bibfnamefont{R.}~\bibnamefont{Shaham}},
  \bibinfo{author}{\bibfnamefont{E.~S.} \bibnamefont{Polzik}},
  \bibnamefont{and}
  \bibinfo{author}{\bibfnamefont{O.}~\bibnamefont{Firstenberg}},
  \bibinfo{journal}{Phys. Rev. Lett.} \textbf{\bibinfo{volume}{124}},
  \bibinfo{pages}{043602} (\bibinfo{year}{2020}).

\bibitem[{\citenamefont{Shen et~al.}(2020)\citenamefont{Shen, Kuhn, Baxter,
  Dalgleish, Li, Lu, Ortiz, Plomp, Parnell, Pynn et~al.}}]{Shen2020}
\bibinfo{author}{\bibfnamefont{J.}~\bibnamefont{Shen}},
  \bibinfo{author}{\bibfnamefont{S.}~\bibnamefont{Kuhn}},
  \bibinfo{author}{\bibfnamefont{D.}~\bibnamefont{Baxter}},
  \bibinfo{author}{\bibfnamefont{R.}~\bibnamefont{Dalgleish}},
  \bibinfo{author}{\bibfnamefont{F.}~\bibnamefont{Li}},
  \bibinfo{author}{\bibfnamefont{S.}~\bibnamefont{Lu}},
  \bibinfo{author}{\bibfnamefont{G.}~\bibnamefont{Ortiz}},
  \bibinfo{author}{\bibfnamefont{J.}~\bibnamefont{Plomp}},
  \bibinfo{author}{\bibfnamefont{S.}~\bibnamefont{Parnell}},
  \bibinfo{author}{\bibfnamefont{R.}~\bibnamefont{Pynn}}, \bibnamefont{et~al.},
  \bibinfo{journal}{Nature Comm.} \textbf{\bibinfo{volume}{11}},
  \bibinfo{pages}{930} (\bibinfo{year}{2020}).

\bibitem[{\citenamefont{Lu et~al.}(2020)\citenamefont{Lu, Irfan, Shen, Kuhn,
  Snow, Baxter, Pynn, and Ortiz}}]{Lu2020}
\bibinfo{author}{\bibfnamefont{S.}~\bibnamefont{Lu}},
  \bibinfo{author}{\bibfnamefont{A.~M.} \bibnamefont{Irfan}},
  \bibinfo{author}{\bibfnamefont{J.}~\bibnamefont{Shen}},
  \bibinfo{author}{\bibfnamefont{S.~J.} \bibnamefont{Kuhn}},
  \bibinfo{author}{\bibfnamefont{W.~M.} \bibnamefont{Snow}},
  \bibinfo{author}{\bibfnamefont{D.~V.} \bibnamefont{Baxter}},
  \bibinfo{author}{\bibfnamefont{R.}~\bibnamefont{Pynn}}, \bibnamefont{and}
  \bibinfo{author}{\bibfnamefont{G.}~\bibnamefont{Ortiz}},
  \bibinfo{journal}{Phys. Rev. A} \textbf{\bibinfo{volume}{101}},
  \bibinfo{pages}{042318} (\bibinfo{year}{2020}).

\bibitem[{\citenamefont{Kuhn et~al.}(2021)\citenamefont{Kuhn, McKay, Shen,
  Geerits, Dalgliesh, Dees, Irfan, Li, Lu, Vangelista et~al.}}]{Kuhn2021}
\bibinfo{author}{\bibfnamefont{S.~J.} \bibnamefont{Kuhn}},
  \bibinfo{author}{\bibfnamefont{S.}~\bibnamefont{McKay}},
  \bibinfo{author}{\bibfnamefont{J.}~\bibnamefont{Shen}},
  \bibinfo{author}{\bibfnamefont{N.}~\bibnamefont{Geerits}},
  \bibinfo{author}{\bibfnamefont{R.}~\bibnamefont{Dalgliesh}},
  \bibinfo{author}{\bibfnamefont{E.}~\bibnamefont{Dees}},
  \bibinfo{author}{\bibfnamefont{A.~A.~M.} \bibnamefont{Irfan}},
  \bibinfo{author}{\bibfnamefont{F.}~\bibnamefont{Li}},
  \bibinfo{author}{\bibfnamefont{S.}~\bibnamefont{Lu}},
  \bibinfo{author}{\bibfnamefont{V.}~\bibnamefont{Vangelista}},
  \bibnamefont{et~al.}, \bibinfo{journal}{Phys. Rev. Research.}
  \textbf{\bibinfo{volume}{3}}, \bibinfo{pages}{023277} (\bibinfo{year}{2021}).

\bibitem[{\citenamefont{Sakharov}(1967)}]{Sakharov1967}
\bibinfo{author}{\bibfnamefont{A.~D.} \bibnamefont{Sakharov}},
  \bibinfo{journal}{Pisma Zh. Eksp. Teor. Fiz} \textbf{\bibinfo{volume}{5}},
  \bibinfo{pages}{32} (\bibinfo{year}{1967}).

\bibitem[{\citenamefont{Mitchell et~al.}(1999)\citenamefont{Mitchell, Bowman,
  and Weidenm\"uller}}]{Mitchell1999}
\bibinfo{author}{\bibfnamefont{G.~E.} \bibnamefont{Mitchell}},
  \bibinfo{author}{\bibfnamefont{J.~D.} \bibnamefont{Bowman}},
  \bibnamefont{and} \bibinfo{author}{\bibfnamefont{H.~A.}
  \bibnamefont{Weidenm\"uller}}, \bibinfo{journal}{Rev. Mod. Phys.}
  \textbf{\bibinfo{volume}{71}}, \bibinfo{pages}{445} (\bibinfo{year}{1999}).

\bibitem[{\citenamefont{Mitchell et~al.}(2001)\citenamefont{Mitchell, Bowman,
  Penttila, and Sharapov}}]{Mitchell2001}
\bibinfo{author}{\bibfnamefont{G.}~\bibnamefont{Mitchell}},
  \bibinfo{author}{\bibfnamefont{J.}~\bibnamefont{Bowman}},
  \bibinfo{author}{\bibfnamefont{S.}~\bibnamefont{Penttila}}, \bibnamefont{and}
  \bibinfo{author}{\bibfnamefont{E.}~\bibnamefont{Sharapov}},
  \bibinfo{journal}{Physics Reports} \textbf{\bibinfo{volume}{354}},
  \bibinfo{pages}{157} (\bibinfo{year}{2001}).

\bibitem[{\citenamefont{Sushkov and Flambaum}(1980)}]{Sushkov1980}
\bibinfo{author}{\bibfnamefont{O.~P.} \bibnamefont{Sushkov}} \bibnamefont{and}
  \bibinfo{author}{\bibfnamefont{V.~V.} \bibnamefont{Flambaum}},
  \bibinfo{journal}{JETP Lett.} \textbf{\bibinfo{volume}{32}},
  \bibinfo{pages}{352} (\bibinfo{year}{1980}).

\bibitem[{\citenamefont{Sushkov and Flambaum}(1982)}]{Sushkov1982}
\bibinfo{author}{\bibfnamefont{O.~P.} \bibnamefont{Sushkov}} \bibnamefont{and}
  \bibinfo{author}{\bibfnamefont{V.~V.} \bibnamefont{Flambaum}},
  \bibinfo{journal}{Soviet Phys. Usp.} \textbf{\bibinfo{volume}{25}},
  \bibinfo{pages}{1} (\bibinfo{year}{1982}).

\bibitem[{\citenamefont{Bunakov and Gudkov}(1982)}]{Bunakov1982}
\bibinfo{author}{\bibfnamefont{V.~R.} \bibnamefont{Bunakov}} \bibnamefont{and}
  \bibinfo{author}{\bibfnamefont{V.}~\bibnamefont{Gudkov}},
  \bibinfo{journal}{Zeitschrift für Physik A Atoms and Nuclei}
  \textbf{\bibinfo{volume}{308}}, \bibinfo{pages}{363–364}
  (\bibinfo{year}{1982}).

\bibitem[{\citenamefont{Gudkov}(1991)}]{Gudkov1991}
\bibinfo{author}{\bibfnamefont{V.}~\bibnamefont{Gudkov}},
  \bibinfo{journal}{Nuclear Physics A} \textbf{\bibinfo{volume}{524}},
  \bibinfo{pages}{668} (\bibinfo{year}{1991}).

\bibitem[{\citenamefont{Fadeev and Flambaum}(2019)}]{Fadeev2019}
\bibinfo{author}{\bibfnamefont{P.}~\bibnamefont{Fadeev}} \bibnamefont{and}
  \bibinfo{author}{\bibfnamefont{V.~V.} \bibnamefont{Flambaum}},
  \bibinfo{journal}{Phys. Rev. C} \textbf{\bibinfo{volume}{100}},
  \bibinfo{pages}{015504} (\bibinfo{year}{2019}).

\bibitem[{\citenamefont{Bowman and Gudkov}(2014)}]{Bowman2014}
\bibinfo{author}{\bibfnamefont{J.~D.} \bibnamefont{Bowman}} \bibnamefont{and}
  \bibinfo{author}{\bibfnamefont{V.}~\bibnamefont{Gudkov}},
  \bibinfo{journal}{Phys. Rev. C} \textbf{\bibinfo{volume}{90}},
  \bibinfo{pages}{065503} (\bibinfo{year}{2014}).

\bibitem[{\citenamefont{Dzuba et~al.}(2018)\citenamefont{Dzuba, Flambaum,
  Samsonov, and Stadnik}}]{Dzuba2018}
\bibinfo{author}{\bibfnamefont{V.~A.} \bibnamefont{Dzuba}},
  \bibinfo{author}{\bibfnamefont{V.~V.} \bibnamefont{Flambaum}},
  \bibinfo{author}{\bibfnamefont{I.~B.} \bibnamefont{Samsonov}},
  \bibnamefont{and} \bibinfo{author}{\bibfnamefont{Y.~V.}
  \bibnamefont{Stadnik}}, \bibinfo{journal}{Phys. Rev. D}
  \textbf{\bibinfo{volume}{98}}, \bibinfo{pages}{035048}
  (\bibinfo{year}{2018}).

\bibitem[{\citenamefont{Mantry et~al.}(2014)\citenamefont{Mantry, Pitschmann,
  and Ramsey-Musolf}}]{Mantry2014}
\bibinfo{author}{\bibfnamefont{S.}~\bibnamefont{Mantry}},
  \bibinfo{author}{\bibfnamefont{M.}~\bibnamefont{Pitschmann}},
  \bibnamefont{and} \bibinfo{author}{\bibfnamefont{M.~J.}
  \bibnamefont{Ramsey-Musolf}}, \bibinfo{journal}{Phys. Rev. D}
  \textbf{\bibinfo{volume}{90}}, \bibinfo{pages}{054016}
  (\bibinfo{year}{2014}).

\end{thebibliography}


\end{document}